\documentclass[journal]{IEEEtran}

\usepackage[utf8]{inputenc}
\usepackage{amsmath}
\usepackage{amssymb}
\usepackage{amsfonts}
\usepackage{mathtools}
\usepackage{bm}
\usepackage{graphicx}
\usepackage{booktabs}
\usepackage{multirow}
\usepackage{array}
\usepackage{tabularx}
\usepackage{algorithm}
\usepackage{algpseudocode}
\usepackage{url}
\usepackage{tcolorbox}
\usepackage{fvextra}
\usepackage{xcolor}
\usepackage{cite}
\usepackage{balance}

\usepackage[colorlinks=true,
            linkcolor=blue,
            citecolor=blue,
            urlcolor=blue,
            filecolor=blue]{hyperref}

\begin{document}

\title{SGTO-MAS: Secure Gorilla Troops Optimization for Multi-Agent LLM Systems}
\author{
Saeid~Jamshidi

\thanks{
S. Jamshidi is with the Department of Computer and Software Engineering,
Polytechnique Montréal, Montréal, QC H3T 1J4, Canada
(e-mail: saeid.jamshidi@polymtl.ca).
}%
}
\maketitle
\begin{abstract}
Multi-agent large language model (LLM) systems have emerged as a powerful paradigm for complex reasoning and decision-making. However, coordinating multiple agents introduces significant challenges, including error propagation, elevated security risk, and inefficient resource utilization. Existing approaches primarily rely on heuristic and static strategies and lack a principled mechanism to balance performance, security, and computational cost. In this paper, we formulate multi-agent LLM coordination as a constrained optimization problem and propose a security-aware approach for adaptive agent selection. The method integrates trust modeling, risk-aware evaluation, and collective intelligence into a unified optimization formulation. To solve this problem efficiently, we leverage a swarm-intelligence strategy inspired by Gorilla Troops Optimization (GTO), enabling adaptive coordination under varying threat conditions. Controlled experiments over 500 independent runs demonstrate the effectiveness of the proposed approach. The system achieves a stable average performance score of 0.5281, with stable consensus (0.8764) and controlled risk (0.3000), while maintaining compact agent subsets with an average of 4.04 selected agents. The optimization process converges efficiently with an average runtime of 24.09 seconds per run and low score variability (standard deviation = 0.0173). Robustness analysis further indicates graceful degradation under perturbations, with performance drops limited to 2.5\% under agent removal and 5.3\% under consensus disruption. The results show that effective multi-agent coordination can be achieved through structured optimization that jointly manages performance, security, and efficiency. The proposed approach provides a practical, security-aware solution for coordinating multi-agent LLM systems in complex, adversarial settings.
\end{abstract}

\begin{IEEEkeywords}
Large Language Models, Multi-Agent Systems, Swarm Intelligence, Gorilla Troops Optimization
\end{IEEEkeywords}
\maketitle

\section{Introduction}
\label{sec:introduction}
Large Language Models (LLMs) have become central to advanced intelligent systems, providing strong capabilities in reasoning, planning, and complex decision-making~\cite{huang2025foundation,valmeekam2023planning}. Recent research extends these models to multi-agent configurations, in which multiple LLM-based agents collaborate on complex tasks through distributed reasoning and coordinated interaction~\cite{ref1,ref2}. Multi-agent LLM systems support task decomposition, parallel execution, and scalable problem solving, addressing key limitations of single-agent architectures\cite{yang2026agentnet}\cite{shaikh2025llm}.\\
Despite these advantages, multi-agent integration introduces substantial challenges. Errors and uncertainties may propagate through shared context, intermediate reasoning steps, and tool interactions~\cite{li2024survey,shaikh2025llm}. Empirical studies show that multi-agent LLMs are vulnerable to hallucination propagation and coordination inconsistencies~\cite{ref3}, with uncertainty arising from inter-agent dependencies, dynamic runtime behavior, and complex interaction patterns~\cite{ref4}. Such vulnerabilities increase system risk and complicate reliable decision-making.
Existing approaches emphasize collaboration protocols, communication strategies, and architectural configurations~\cite{ref1}. These methods improve coordination efficiency but rely heavily on static and heuristic agent-selection policies, with limited mechanisms to dynamically balance decision quality, security risk, and computational cost. Application-driven implementations, including multi-agent LLMs for cybersecurity and fuzz testing, demonstrate strong empirical results~\cite{ref5}, but they do not sufficiently address adaptive agent selection under adversarial conditions. Existing systems often rely on predefined role assignments, fixed interaction patterns, and broad tool-enabled execution, increasing exposure to vulnerabilities such as prompt injection, unsafe tool execution, and data leakage~\cite{wu2024autogen,li2023camel,ferrag2025threats}.\\
To address these gaps, this study formulates multi-agent LLM coordination as a constrained optimization problem. SGTO-MAS introduces a security-aware optimization mechanism that dynamically selects agents based on trust metrics, query-specific risk estimates, and collective-intelligence signals. This design supports high-quality decision-making by jointly balancing performance, security resilience, and computational efficiency. The optimization process incorporates principles from Gorilla Troops Optimization (GTO)~\cite{van2026artificial}, using swarm-intelligence dynamics to guide adaptive agent selection. Coupling GTO with trust-aware modeling and query-sensitive risk assessment enables robust coordination of heterogeneous agents under varying threat conditions. The evaluation shows that SGTO-MAS produces stable outputs with controlled risk and efficient execution. The results further indicate that effective coordination emerges from structured interactions, controlled diversity, and trust-weighted aggregation rather than indiscriminate scaling of agent participation. The primary contributions of this study are:
\begin{itemize}
    \item \textbf{Optimization-driven multi-agent coordination:} This study formulates multi-agent coordination as a constrained optimization problem, enabling dynamic agent selection beyond static heuristics.
    \item \textbf{Security-aware coordination with trust and risk modeling:} The proposed formulation integrates trust modeling, query-specific risk assessment, and collective intelligence to support robust and secure decision-making.
    \item \textbf{Swarm-intelligence integration via GTO:} A GTO-inspired optimization strategy is introduced for agent selection and coordination in multi-agent LLM systems, connecting swarm-intelligence principles with agentic AI.
\end{itemize}

The remainder of this paper is structured as follows. Section~\ref{sec:related_work} reviews related work, Section~\ref{sec:method} presents the SGTO-MAS methodology, Section~\ref{sec:threat_model} defines the threat model, and Section~\ref{sec:experimental_setup} describes the experimental setup. Section~\ref{sec:results} reports the experimental results, Section~\ref{sec:discussion} discusses the main findings, Section~\ref{sec:limitations_future} outlines limitations and future work, and Section~\ref{sec:conclusion} concludes the paper.

\section{Related Work}
\label{sec:related_work}
This section reviews prior work on multi-agent LLM coordination, collaborative reasoning, uncertainty, and security-awareness.

\subsection{LLM-Based Multi-Agent Systems}
The rapid advancement of LLMs has enabled multi-agent systems in which agents collaborate to solve complex tasks through distributed reasoning and parallel execution \cite{ref1,ref2}. These systems support task decomposition, enhanced scalability, and cross-agent validation \cite{ref6}, establishing the importance of multi-agent collaboration. Existing studies primarily focus on system design and collaboration mechanisms, including agent roles, communication protocols, and coordination strategies, but lack adaptive agent selection.

\subsection{Agent Interaction Mechanisms}
Prior frameworks such as AutoGen and CAMEL demonstrate structured multi-agent interaction, role-based collaboration, and dynamic task delegation \cite{ref7,ref8}. These methods improve cooperative behaviors and task-oriented conversations. Nevertheless, they rely on predefined agent roles and static configurations, without mechanisms to dynamically optimize agent participation based on task requirements and runtime system conditions.

\subsection{Collaborative Reasoning and Collective Intelligence}
Multi-agent debate and iterative reasoning approaches indicate that agent interaction can improve factual accuracy and reasoning performance \cite{ref9}. Diversity and complementarity among agents enable emergent collective intelligence. Existing approaches, however, assume fixed agent sets and do not explicitly optimize group composition, limiting adaptability and efficiency in dynamic scenarios.

\subsection{Uncertainty, Reliability, and Security}
Multi-agent LLM systems introduce significant uncertainty due to inter-agent dependencies and iterative interactions \cite{ref4}. Hallucination propagation and inconsistent reasoning have been identified as critical vulnerabilities \cite{ref3}. These challenges increase system susceptibility, particularly in adversarial environments where malicious inputs exploit coordination weaknesses. Most existing work treats risk as a secondary concern and does not integrate it into coordination mechanisms.

\subsection{Application-Oriented Systems}
Real-world applications demonstrate the potential of multi-agent LLMs. For instance, MALF leverages agent collaboration for cybersecurity tasks, improving vulnerability detection \cite{ref5}. Industrial deployments demonstrate the effectiveness of automation, decision support, and large-scale AI services \cite{ref10}. Despite strong empirical results, these implementations typically rely on static configurations and lack adaptivity.\\

Existing literature demonstrates significant progress in collaborative reasoning, architecture, and application deployment. Nevertheless, current systems still depend largely on static coordination and heuristic mechanisms. Adaptive agent selection, trust modeling, risk-aware decision-making, and system-level security remain insufficiently addressed. To address these gaps, this work introduces an optimization-driven approach for multi-agent coordination, formulating agent selection as a constrained optimization problem. The proposed method integrates trust, risk, and collective intelligence, leveraging swarm-inspired GTO, to enable dynamic, secure, and efficient heterogeneous LLM agents. This design supports a principled balance among performance, security resilience, and computational cost.

\section{Methodology}
\label{sec:method}
\begin{figure*}[!t]
\centering
\includegraphics[width=0.90\textwidth]{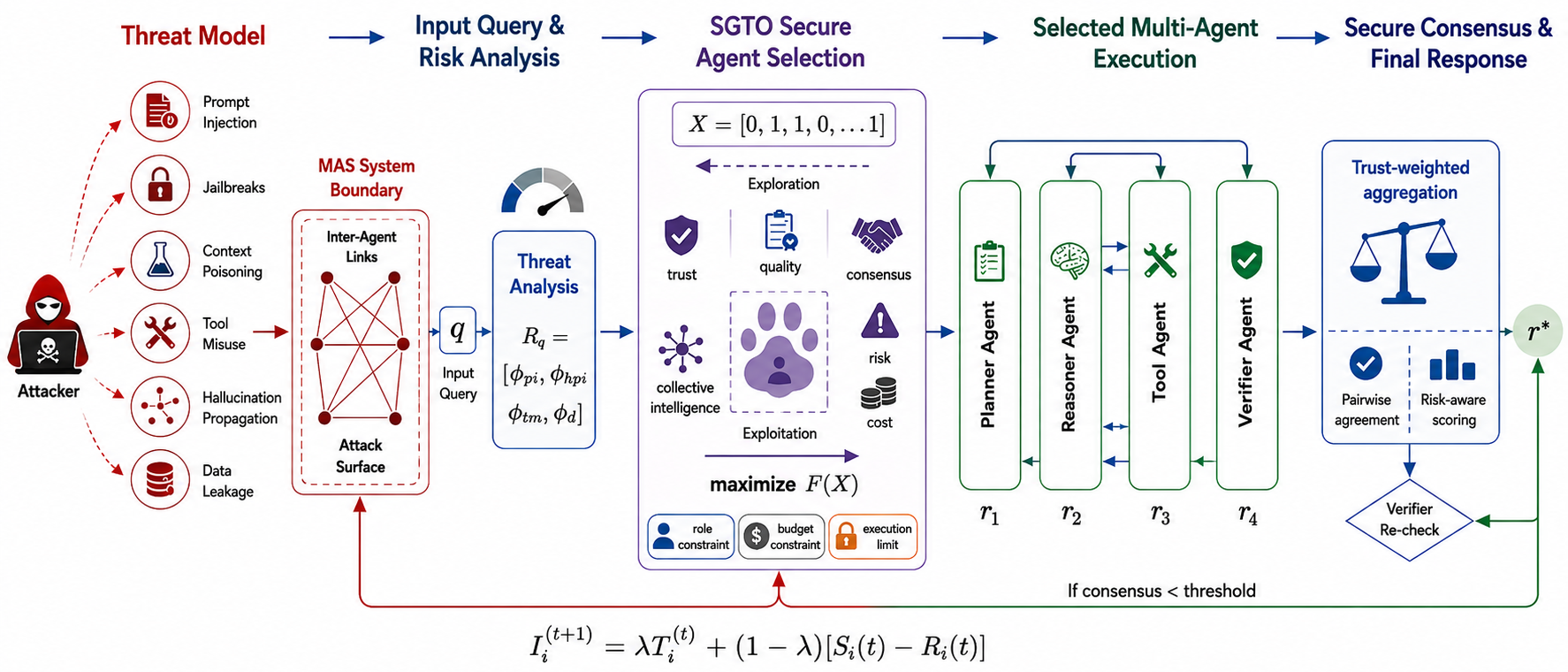}
\caption{Overview of the proposed SGTO-MAS approach. The system performs threat-aware query analysis, followed by optimization-based agent selection using a GTO-inspired mechanism, multi-agent execution, and trust-weighted consensus aggregation. Dynamic trust updates continuously refine agent reliability over time.}
\label{fig:sgto_architecture}
\end{figure*}

This section presents the proposed approach for secure coordination in multi-agent LLM systems. The goal is to enable reliable collaboration among heterogeneous agents by mitigating system-level risks, including prompt injection, hallucination propagation, unsafe tool usage, and data leakage. Agent coordination is formulated as a security-aware optimization problem, in which the system dynamically selects an effective subset of agents based on trust, query-dependent risk, and expected task contribution. The proposed method operates through a structured pipeline, illustrated in Figure~\ref{fig:sgto_architecture}, comprising four stages: threat-aware query analysis, optimization-based agent selection, multi-agent execution, and trust-weighted consensus aggregation with dynamic trust updates. For a given input query, the system first evaluates its threat profile, capturing risks that may impact downstream reasoning. Given this threat context, the optimization module identifies the most suitable subset of agents by balancing response quality and security exposure. Selected agents then collaboratively generate outputs, which are aggregated through a trust-aware consensus mechanism. Agent trust scores are updated after each execution to support adaptive and robust coordination over time.
Multi-agent LLM systems support complex problem-solving by enabling specialized reasoners, planners, tools, and verifiers to interact coherently. Although this distributed design enhances scalability and task decomposition, it also enlarges the attack surface. In contrast to single-agent systems, errors and adversarial behaviors may propagate across agents, compromising downstream decisions. Consequently, the central challenge is to coordinate agents to maximize output quality and minimize risk exposure. Formally, let the set of available agents be defined as
\begin{equation}
A = \{a_1, a_2, \dots, a_N\}.
\end{equation}
Each agent $a_i$ produces a response
\begin{equation}
r_i = f_i(q, \mathcal{M}_i, \mathcal{T}_i),
\end{equation}
where $q$ denotes the input query, $\mathcal{M}_i$ represents internal agent context, and $\mathcal{T}_i$ denotes accessible tools. Activating all agents increases computational cost and risk exposure, whereas selecting too few agents may reduce robustness. To address this trade-off, the optimal subset of agents is defined as
\begin{equation}
X^\star = \arg \max_{X \in \{0,1\}^N} F(X \mid q, R_q, T),
\end{equation}
where $R_q$ is the query-dependent threat vector and $T$ represents the trust-state vector of agents. The system state at iteration $t$ is represented as
\begin{equation}
\mathcal{S}(t) = \big(q, R_q, T(t), \Pi(t), \mathcal{C}(t)\big),
\end{equation}
where $T(t)$ captures trust dynamics, $\Pi(t)$ denotes agent-role assignments, and $\mathcal{C}(t)$ records historical coordination context. To efficiently solve this combinatorial optimization problem, agent selection is modeled using a swarm-intelligence strategy inspired by GTO. Each candidate solution corresponds to a subset of agents, and the optimization process seeks configurations that maximize a security-aware objective function accounting for trust, risk exposure, response quality, and coordination effectiveness. By integrating threat-aware modeling, swarm-based optimization, and trust-driven aggregation, the method supports adaptive, secure, and scalable coordination across heterogeneous LLM agents.

\subsection{System Architecture}
The proposed SGTO-MAS architecture comprises four interconnected components that enable secure, adaptive coordination in multi-agent LLM systems: threat analysis, optimization-based agent selection, multi-agent execution, and trust-weighted consensus aggregation. These components exchange shared signals, including threat levels, trust scores, and response consistency. The threat analysis module evaluates an input query $q$ for adversarial patterns, including prompt injection, unsafe instructions, hallucination propagation, and data leakage. The evaluation produces a threat vector defined as
\begin{equation}
R_q = (\phi_{pi}, \phi_{hp}, \phi_{tm}, \phi_{dl}),
\end{equation}
where $\phi_{pi}$ denotes the estimated risk of prompt injection, $\phi_{hp}$ represents the likelihood of hallucination propagation across agents, $\phi_{tm}$ quantifies the probability of unsafe tool misuse, and $\phi_{dl}$ indicates the risk of data leakage. The optimization module selects an agent subset that balances trust, role suitability, and risk exposure. Each candidate solution is encoded as a binary vector
\begin{equation}
X = (x_1, x_2, \dots, x_N), \quad x_i =
\begin{cases}
1, & \text{if agent } a_i \text{ is selected},\\
0, & \text{if agent } a_i \text{ is not selected}.
\end{cases}
\end{equation}
The set of active agents for the current query is
\begin{equation}
A_X = \{a_i \in A \mid x_i = 1\},
\end{equation}
with each active agent $a_i$ generating a response $r_i$, forming the response set
\begin{equation}
R = \{r_1, r_2, \dots, r_K\}, \quad K = \sum_{i=1}^{N} x_i,
\end{equation}
where $K$ is the number of selected agents. Agent outputs are aggregated to quantify agreement using the metric
\begin{equation}
C = \frac{1}{K(K-1)} \sum_{i \ne j} sim(r_i,r_j),
\end{equation}
where $sim(r_i,r_j)$ measures response similarity between agents $i$ and $j$, capturing consensus among active agents. The final system output is computed as a function of the query $q$, threat vector $R_q$, optimized agent selection $X^\star$, trust scores $T$, and agent responses:
\begin{equation}
r^\star = \mathcal{G}\big(q, R_q, X^\star, T, \{r_i\}_{i=1}^{K}\big),
\end{equation}
where $\mathcal{G}(\cdot)$ integrates adaptive agent selection, risk awareness, trust weighting, and consensus aggregation to produce the final response. This formulation supports adaptive agent selection based on threat and trust signals, reduces exposure to system-level risks through selective activation, and improves response reliability through trust-weighted consensus.

\subsection{Threat Quantification and Query Risk Modeling}
The proposed coordination mechanism adapts to query-specific threats, enabling differentiated responses to benign and adversarial inputs. Each query $q$ is transformed into a structured threat vector that captures risks across multiple dimensions. The normalized threat vector is defined as
\begin{equation}
R_q = [\phi_{pi}, \phi_{hp}, \phi_{tm}, \phi_{dl}]^\top,
\end{equation}
where each component $\phi_j \in [0,1]$ represents the estimated risk of prompt injection ($\phi_{pi}$), hallucination propagation ($\phi_{hp}$), unsafe tool usage ($\phi_{tm}$), and data leakage ($\phi_{dl}$). In the experimental implementation, these values are treated as normalized query-level risk indicators derived from the assigned threat category and used consistently across optimization, exposure estimation, and risk logging. An aggregate threat score for the query is computed as
\begin{equation}
\Omega_q = \sum_{j=1}^{4} \omega_j \phi_j, \quad \text{subject to } \sum_{j=1}^{4} \omega_j = 1,
\end{equation}
where $\omega_j$ controls the relative importance of each threat component. The value of $\Omega_q$ guides agent selection: low-risk queries permit broader participation, whereas high-risk queries favor conservative subsets with stronger trust scores. To incorporate query-dependent risk at the agent level, each agent $a_i$ is assigned an exposure score
\begin{equation}
E_i(q) = \sum_{j=1}^{4} \mu_j \phi_j v_i^j,
\end{equation}
where $v_i^j$ quantifies the vulnerability of agent $a_i$ to threat type $j$, and $\mu_j$ denotes the corresponding threat weight. The security suitability score for agent $a_i$ under query $q$ is then defined as
\begin{equation}
\Lambda_i(q) = \frac{T_i}{1 + E_i(q)},
\end{equation}
balancing trust $T_i$ and risk exposure. Higher values indicate agents with stronger reliability under the current threat conditions. To improve temporal stability, a smoothed threat estimate is maintained as
\begin{equation}
\bar{R}_q(t) = \chi \bar{R}_q(t-1) + (1-\chi) R_q(t),
\end{equation}
where $\chi \in [0,1]$ controls the impact of past observations. This smoothing step reduces sensitivity to transient fluctuations and noisy query-level risk estimates.

\subsection{Security-Aware Fitness Function}
The fitness function evaluates each candidate agent subset by jointly considering security, reliability, consensus, and computational cost. In contrast to single-metric objectives that emphasize task accuracy alone, this formulation balances decision performance with risk exposure. For a selection vector $X$, the fitness is defined as
\begin{equation}
\begin{aligned}
F(X) =\;& \alpha ASB(X) + \beta TR(X) + \gamma CR(X) + \eta QA(X) \\
&- \delta HR(X) - \epsilon TMR(X) - \zeta DL(X) - \kappa LC(X),
\end{aligned}
\end{equation}
where $\alpha,\beta,\gamma,\eta$ weight desirable properties and $\delta,\epsilon,\zeta,\kappa$ weight unsafe behaviors and computational cost. Higher $F(X)$ values indicate agent subsets with stronger security-performance trade-offs. The trust component is computed as
\begin{equation}
TR(X) = \frac{1}{K} \sum_{i=1}^{N} x_i T_i,
\end{equation}
where $T_i$ denotes the trust score of agent $a_i$, and $K = \sum_{i=1}^{N} x_i$ is the number of selected agents. Historical agreement among agents is quantified by
\begin{equation}
CR(X) = \frac{1}{K(K-1)} \sum_{i \ne j} x_i x_j \Gamma_{ij},
\end{equation}
where $\Gamma_{ij}$ represents past compatibility between agents $i$ and $j$. Agent quality over prior tasks is included as
\begin{equation}
QA(X) = \frac{1}{K} \sum_{i=1}^{N} x_i Q_i,
\end{equation}
where $Q_i$ is the historical quality score of agent $a_i$. Risks are incorporated as
\begin{equation}
\begin{aligned}
HR(X) &= \frac{1}{K} \sum_{i=1}^{N} x_i H_i, \\
TMR(X) &= \frac{1}{K} \sum_{i=1}^{N} x_i U_i, \\
DL(X) &= \frac{1}{K} \sum_{i=1}^{N} x_i L_i.
\end{aligned}
\end{equation}
where $H_i$, $U_i$, and $L_i$ denote hallucination, unsafe tool usage, and data leakage risks, respectively. Computational cost is represented as
\begin{equation}
LC(X) = \sum_{i=1}^{N} x_i c_i,
\end{equation}
where $c_i$ corresponds to the resource cost of agent $a_i$. To account for query-specific robustness, the attack suppression benefit is defined as
\begin{equation}
ASB(X) = \frac{1}{K} \sum_{i=1}^{N} x_i (1 - E_i(q)) T_i,
\end{equation}
rewarding agents that are both trustworthy and less vulnerable under the current query $q$. This formulation separates beneficial and penalized contributions:
\begin{equation}
F(X) = U_{\text{secure}}(X) - \mathcal{R}_{\text{unsafe}}(X),
\end{equation}
with
\begin{equation}
\begin{aligned}
U_{\text{secure}}(X) &= \alpha ASB + \beta TR + \gamma CR + \eta QA, \\
\mathcal{R}_{\text{unsafe}}(X) &= \delta HR + \epsilon TMR + \zeta DL + \kappa LC.
\end{aligned}
\end{equation}
This decomposition improves interpretability by distinguishing desirable properties from penalized risks. A sparsity penalty is added to constrain unnecessary agent activation:
\begin{equation}
F_{\text{aug}}(X) = F(X) - \psi \|X\|_0,
\end{equation}
encouraging compact, reliable, and low-risk agent subsets without sacrificing decision quality and system efficiency.

\subsection{Collective Intelligence and Robust Consensus Modeling}
Effective multi-agent LLM coordination requires preserving Collective Intelligence (CI), defined as the emergent reasoning capability produced when heterogeneous agents collaborate under secure constraints. Multi-agent reasoning benefits from diverse perspectives, role specialization, and distributed access to evidence. Excessive redundancy among agents reduces individual contributions, whereas excessive disagreement can degrade reliability. SGTO-MAS balances diversity and consensus to support secure collective intelligence.
Let $X=(x_1,x_2,\dots,x_N)$ denote the binary selection vector, where $x_i=1$ indicates an active agent $a_i$. The collective intelligence score is defined as
\begin{equation}
\begin{aligned}
CI(X) =\;& \lambda_1 DIV(X) + \lambda_2 COMP(X) \\
&+ \lambda_3 SYN(X) + \lambda_4 C_w(X).
\end{aligned}
\end{equation}
where $DIV(X)$ quantifies cognitive diversity, $COMP(X)$ measures agent complementarity, $SYN(X)$ captures collaborative synergy, and $C_w(X)$ computes trust-weighted consensus. The coefficients satisfy
\begin{equation}
\lambda_1 + \lambda_2 + \lambda_3 + \lambda_4 = 1,
\end{equation}
ensuring a normalized combination of diversity, complementarity, synergy, and consensus contributions.
Cognitive diversity is defined as
\begin{equation}
DIV(X) = \frac{1}{K(K-1)} \sum_{i \ne j} x_i x_j d_{ij},
\end{equation}
where $d_{ij} \in [0,1]$ represents pairwise heterogeneity between agents $a_i$ and $a_j$, and $K = \sum_{i=1}^{N} x_i$. A higher $DIV(X)$ indicates more heterogeneous reasoning, thereby improving coverage and resilience.
Agent complementarity measures the extent to which agents compensate for each other:
\begin{equation}
COMP(X) = \frac{1}{K(K-1)} \sum_{i \ne j} x_i x_j \Delta_{ij},
\end{equation}
where $\Delta_{ij} \in [0,1]$ indicates the complementarity between agents $i$ and $j$, supporting broader reasoning capability.
Collaborative synergy is defined as
\begin{equation}
SYN(X) = \frac{1}{K} \sum_{i=1}^{N} x_i \Omega_i + \frac{1}{K(K-1)} \sum_{i \ne j} x_i x_j \Theta_{ij},
\end{equation}
where $\Omega_i$ is the individual reasoning utility and $\Theta_{ij}$ is the pairwise collaborative gain.
Trust-weighted consensus is defined as
\begin{equation}
C_w(X) = \frac{\sum_{i \ne j} T_i T_j sim(r_i,r_j)}{\sum_{i \ne j} T_i T_j},
\end{equation}
giving stronger impact to agents with higher trust values. To reduce artificial consensus under adversarial conditions, a groupthink penalty is introduced:
\begin{equation}
GT(X) = \max(0, C_w(X)-\tau) \cdot (1-DIV(X)),
\end{equation}
where $\tau$ is the acceptable consensus threshold. This term penalizes highly homogeneous selections with excessive agreement. Functional role balance is modeled as
\begin{equation}
RB(X) = \sum_{r \in \mathcal{R}} \left| \frac{\sum_{i \in r} x_i}{K} - \pi_r \right|,
\end{equation}
where $\pi_r$ denotes the desired proportion of role $r$ in the active subset.
The final optimization objective integrates CI with the security-aware fitness function:
\begin{equation}
\begin{aligned}
F_{\text{total}}(X) =\;& \alpha ASB(X) + \beta TR(X) + \gamma CR(X) + \eta QA(X) \\
&+ \mu CI(X) - \delta HR(X) - \epsilon TMR(X) \\
&- \zeta DL(X) - \kappa LC(X) - \phi GT(X) - \psi RB(X),
\end{aligned}
\end{equation}
where $\mu$ weights collective intelligence, and $\phi$ and $\psi$ penalize groupthink and role imbalance, respectively. This formulation supports selecting agent subsets that balance security, trust, diversity, and collaborative effectiveness.

\subsection{Dynamic Trust Modeling}
SGTO-MAS maintains a dynamic trust score for each agent, representing the system's current estimate of agent reliability and safety based on recent interactions. Let $T_i(t)$ denote the trust score of agent $a_i$ at interaction round $t$. The trust update is defined as
\begin{equation}
T_i(t+1) = \lambda T_i(t) + (1-\lambda) \big(S_i(t) - R_i(t)\big),
\end{equation}
where $\lambda \in [0,1]$ is a temporal smoothing coefficient, $S_i(t)$ is the security compliance score, and $R_i(t)$ is the cumulative risk penalty. The compliance score assesses whether an agent adheres to policy constraints, provides valid reasoning, respects tool boundaries, and avoids unsafe disclosures. The risk penalty captures observed negative behaviors, including hallucination, unsafe tool usage, data leakage, and susceptibility to malicious instructions:
\begin{equation}
R_i(t) = \nu_1 h_i(t) + \nu_2 u_i(t) + \nu_3 d_i(t) + \nu_4 m_i(t),
\end{equation}
where $h_i(t)$, $u_i(t)$, $d_i(t)$, and $m_i(t)$ quantify hallucination events, unsafe tool interactions, data leakage, and malicious-instruction vulnerability, respectively. The parameters $\nu_1,\nu_2,\nu_3,\nu_4$ weight the severity of each risk type. Trust values are bounded to maintain numerical stability and interpretability:
\begin{equation}
T_i(t+1) \leftarrow \min\{T_{\max}, \max\{T_{\min}, T_i(t+1)\}\},
\end{equation}
with $T_{\min}=0$ and $T_{\max}=1$ in the experimental implementation. A confidence-corrected trust score adjusts for the number of observed interactions:
\begin{equation}
\widetilde{T}_i(t) = T_i(t)\left(1 - e^{-\frac{n_i(t)}{\varsigma}}\right),
\end{equation}
where $n_i(t)$ is the cumulative number of interactions for agent $a_i$, and $\varsigma$ is a scaling constant. This correction reduces overconfident trust assignments for newly introduced agents with limited interaction history. For role-specific adaptation, let $\varrho_i$ denote the functional role of agent $a_i$. The compliance score is decomposed as
\begin{equation}
S_i(t) = \sum_{m=1}^{M_r} \pi_m^{(\varrho_i)} s_{im}(t),
\end{equation}
where $s_{im}(t)$ are role-specific compliance indicators and $\pi_m^{(\varrho_i)}$ are role-dependent importance weights. This extension captures distinct safety expectations among reasoners, planners, tool agents, and verifiers, enabling adaptive trust evaluation tailored to each agent's functionality.

\subsection{Binary Agent Encoding and Feasibility Constraints}
Although SGTO-MAS employs a swarm-based optimization strategy, agent selection is inherently discrete: each agent is either included in the current subset or excluded. Therefore, the continuous position vectors generated by the Gorilla Troops Optimizer \cite{hussien2024depth} are transformed into binary selection decisions. Let $X_i(t)$ denote the continuous position vector of gorilla $i$ at iteration $t$. Each dimension is binarized via a sigmoid transfer function:
\begin{equation}
\sigma(z) = \frac{1}{1 + e^{-z}},
\end{equation}
and the binary update is
\begin{equation}
x_{ij}(t+1) =
\begin{cases}
1, & \text{if } \sigma(X_{ij}(t+1)) > \theta,\\
0, & \text{otherwise},
\end{cases}
\end{equation}
where $\theta \in (0,1)$ is a threshold, and $x_{ij}$ indicates whether agent $j$ is selected in candidate solution $i$. Candidate subsets must satisfy operational feasibility constraints. At least one reasoning agent and one verifier agent must be selected:
\begin{equation}
\sum_{j \in \mathcal{R}} x_j \ge 1, \quad
\sum_{j \in \mathcal{V}} x_j \ge 1,
\end{equation}
where $\mathcal{R}$ and $\mathcal{V}$ denote the sets of reasoning and verifier agents, respectively. In the experimental configuration, each query passes through a constrained tool-safety gate; therefore, at least one tool-capable agent is selected:
\begin{equation}
\sum_{j \in \mathcal{U}} x_j \ge 1,
\end{equation}
where $\mathcal{U}$ denotes the set of tool-capable agents. Budgetary and security constraints are incorporated to limit computational cost and risk exposure:
\begin{equation}
\sum_{i=1}^{N} x_i c_i \le B_{\max}, \quad
\sum_{i=1}^{N} x_i E_i(q) \le \mathcal{E}_{\max},
\end{equation}
where $B_{\max}$ bounds the latency and computational budget, and $\mathcal{E}_{\max}$ specifies the maximum acceptable aggregate exposure. To maintain valid candidate solutions, a repair operator $\mathcal{R}_{\text{repair}}(\cdot)$ may be applied by deactivating low-utility agents, enabling mandatory-role agents, and projecting the candidate onto the feasible subset space until all constraints are satisfied:
\begin{equation}
\widehat{X} = \mathcal{R}_{\text{repair}}(X).
\end{equation}
This approach preserves the simplicity of swarm update equations and keeps evaluated subsets feasible, security-aware, and operationally functional.

\subsection{SGTO-MAS Optimization Process}
The SGTO-MAS optimization procedure is inspired by the social behavior of gorilla troops, in which group members alternate between exploratory movement and convergence toward dominant leaders. Each gorilla represents a candidate agent-subset configuration, and the population is initialized as
\begin{equation}
P = \{X_1, X_2, \dots, X_M\},
\end{equation}
where $M$ is the population size and each $X_i$ is a candidate selection vector.
During optimization, the search alternates between exploration and exploitation. Exploration preserves diversity and searches broadly across agent-subset configurations:
\begin{equation}
X_i(t+1) = X_i(t) + r \cdot (X_{rand}(t) - X_i(t)),
\end{equation}
where $r$ is a stochastic coefficient and $X_{rand}(t)$ is a randomly selected candidate, promoting stochastic coverage of the search space. Exploitation refines promising solutions by moving candidates toward the best solution:
\begin{equation}
X_i(t+1) = X_i(t) + \beta \cdot (X_{best}(t) - X_i(t)),
\end{equation}
where $\beta$ controls convergence strength and $X_{best}(t)$ is the best solution according to the fitness function at iteration $t$.
To integrate security considerations, SGTO-MAS modulates exploration and exploitation based on the aggregate query risk $\Omega_q$. The exploitation coefficient is adjusted as
\begin{equation}
\beta_q = \beta_0 (1 - \Omega_q),
\end{equation}
where $\beta_0$ is the nominal exploitation factor. High-risk queries induce more conservative movements that favor trusted local refinements, whereas low-risk queries permit broader exploratory shifts. Similarly, the exploration amplitude is adapted as
\begin{equation}
r_q = r_0(1 + \Omega_q),
\end{equation}
providing risk-sensitive control over the search trajectory.
After each update, continuous positions are binarized and projected onto the feasible solution space. Fitness values are recomputed, and the global best candidate is updated when improvement is observed. Population diversity is monitored using
\begin{equation}
\bar{X}(t) = \frac{1}{M} \sum_{i=1}^{M} X_i(t), \quad
D(t) = \frac{1}{M} \sum_{i=1}^{M} \|X_i(t) - \bar{X}(t)\|_2.
\end{equation}
If $D(t)$ falls below a predefined threshold, partial reinitialization and mutation may be applied to reduce premature convergence in the binary subset-selection space.
The optimization can be formally expressed as
\begin{equation}
\max_{X \in \mathcal{X}_{\text{feas}}} F_{\text{aug}}(X),
\end{equation}
where $\mathcal{X}_{\text{feas}}$ denotes the feasible binary search space defined by role, budget, and security constraints. This explicit formulation aligns the method with constrained combinatorial optimization and supports secure, high-quality, and operationally valid agent selections.

\subsection{Secure Consensus and Response Aggregation}
After generating responses, SGTO-MAS applies a security-aware consensus mechanism to determine the final output. This process extends beyond simple averaging by modeling disagreement patterns that may indicate hallucination, adversarial contamination, and unsafe tool use.
Let the set of active responses be $R = \{r_1,\dots,r_K\}$. Pairwise semantic agreement is computed to obtain a consensus score:
\begin{equation}
C = \frac{1}{K(K-1)} \sum_{i \ne j} sim(r_i,r_j).
\end{equation}
To reduce the impact of potentially compromised agents, a trust-weighted consensus score is defined as
\begin{equation}
C_w = \frac{\sum_{i \ne j} T_i T_j \, sim(r_i,r_j)}{\sum_{i \ne j} T_i T_j},
\end{equation}
giving higher impact to agents with stronger trust values.
Each agent's final response score combines trust, agreement, answer quality, and local risk:
\begin{equation}
\Psi_i = \xi_1 T_i + \xi_2 \bar{s}_i + \xi_3 q_i - \xi_4 \ell_i,
\end{equation}
where $\bar{s}_i$ is the mean semantic agreement of $r_i$ with other responses, $q_i$ is the internal quality score, and $\ell_i$ represents estimated local risk. The response with the highest $\Psi_i$ may be selected, and a synthesis of top responses may be generated when explanation and composition are required. The mean agreement for each agent is formalized as
\begin{equation}
\bar{s}_i = \frac{1}{K-1} \sum_{j \ne i} sim(r_i,r_j),
\end{equation}
and the local risk is decomposed into specific threat components:
\begin{equation}
\ell_i = \xi_5 \ell_i^{hall} + \xi_6 \ell_i^{tool} + \xi_7 \ell_i^{leak} + \xi_8 \ell_i^{inj},
\end{equation}
where $\ell_i^{hall}$, $\ell_i^{tool}$, $\ell_i^{leak}$, and $\ell_i^{inj}$ denote risks from hallucination, unsafe tool usage, data leakage, and injection contamination.
When response synthesis is required, the final output is generated as a weighted composition of high-scoring, low-risk fragments:
\begin{equation}
r^\star = \mathcal{H}(\{r_i, \Psi_i\}_{i=1}^{K}),
\end{equation}
where $\mathcal{H}(\cdot)$ is a synthesis operator supporting extractive and generative aggregation. This approach supports reliable and security-aware output generation by combining trust weighting, semantic agreement, response quality, and local risk estimation.

\subsection{Algorithm Description}
Algorithm~\ref{alg:sgto_mas} presents the SGTO-MAS optimization procedure for security-aware multi-agent coordination. The method integrates query-level threat analysis, swarm-based agent subset selection, trust-aware evaluation, and consensus-driven aggregation within a unified optimization loop.
Given an input query $q$, the system first computes the threat vector $R_q$ and the aggregate risk score $\Omega_q$. A population of candidate agent subsets is initialized, with each candidate represented as a binary selection vector. Each candidate is evaluated using a security-aware fitness function that jointly considers trust, response quality, consensus robustness, risk exposure, and computational cost.
The optimization proceeds iteratively using a swarm-intelligence strategy inspired by GTO. Candidate solutions alternate between exploratory updates and exploitation toward the current best solution. The search dynamics are modulated by query risk: high-risk queries induce conservative updates that favor trusted local refinements, whereas low-risk queries allow broader exploration. After each update, continuous positions are binarized, projected onto the feasible solution space, repaired when needed, and re-evaluated. The global best solution $X_{best}$ determines the subset of agents activated for collaborative response generation. Responses are aggregated using a trust-weighted consensus mechanism, and the final output $r^\star$ is selected based on semantic agreement, internal quality, and risk scores. Agent trust scores are updated based on observed behavior, supporting adaptive coordination across successive queries.
\begin{algorithm}
\footnotesize
\caption{SGTO-MAS Security-Aware Multi-Agent Optimization}
\label{alg:sgto_mas}
\begin{algorithmic}[1]
\Require Query $q$, agent set $A=\{a_1,\ldots,a_N\}$, population size $M$, maximum iterations $T$
\Ensure Final response $r^\star$

\State Compute threat vector $R_q = (\phi_{pi},\phi_{hp},\phi_{tm},\phi_{dl})$
\State Compute aggregate risk score $\Omega_q$
\State Initialize population $P=\{X_1,\ldots,X_M\}$
\State Initialize trust scores $T=[T_1,\ldots,T_N]$

\For{each $X_i \in P$}
    \State Convert $X_i$ to a binary agent-selection vector
    \State Apply feasibility constraints
    \State Compute fitness $F(X_i)$
\EndFor

\State Identify best solution $X_{best}$

\For{$t=1$ to $T$}
    \State Compute risk-sensitive coefficient $\beta_q=\beta_0(1-\Omega_q)$

    \For{each $X_i \in P$}
        \If{exploration condition is satisfied}
            \State Select random solution $X_{rand}$
            \State $X_i \leftarrow X_i + r(X_{rand}-X_i)$
        \Else
            \State $X_i \leftarrow X_i + \beta_q(X_{best}-X_i)$
        \EndIf

        \State Apply sigmoid binarization
        \State Repair infeasible solutions
        \State Compute fitness $F(X_i)$
    \EndFor

    \State Update global best solution $X_{best}$
\EndFor

\State Activate selected agents $A_{X_{best}}$
\State Collect responses $R=\{r_1,\ldots,r_K\}$
\State Compute trust-weighted consensus
\State Select final response $r^\star$

\For{each active agent $a_i$}
    \State Compute compliance score $S_i$
    \State Compute risk penalty $R_i$
    \State Update trust score:
    \State $T_i \leftarrow \lambda T_i + (1-\lambda)(S_i - R_i)$
\EndFor

\State \Return $r^\star$
\end{algorithmic}
\end{algorithm}

\section{Threat Model}
\label{sec:threat_model}
Security-aware coordination in multi-agent LLM systems must be evaluated under adversarial conditions that reflect practical attack surfaces. We formalize adversarial capabilities, attack surfaces, and security objectives governing multi-agent coordination, linking them to threat categories such as prompt injection, jailbreak attacks, context poisoning, and tool misuse.

\subsection{System Model}
We consider a multi-agent LLM system composed of heterogeneous agents
\begin{equation}
A = \{a_1, a_2, \dots, a_N\},
\end{equation}
each performing specialized functions such as reasoning, planning, constrained tool interaction, and verification. Agents communicate through intermediate messages during reasoning. Given a query $q$, a subset of agents is activated:
\begin{equation}
A_X = \{a_i \in A \mid x_i = 1\},
\end{equation}
where $X=(x_1, \dots, x_N)$ is a binary selection vector. Each active agent $a_i$ produces a response
\begin{equation}
r_i = f_i(q, m_i, t_i),
\end{equation}
with $m_i$ representing internal context and $t_i$ denoting accessible tools and resources. The final system response is obtained through a trust-aware aggregation function
\begin{equation}
r^\star = \mathcal{G}(r_1, \dots, r_K),
\end{equation}
where $\mathcal{G}(\cdot)$ is the aggregation operator and $K = \sum_{i=1}^{N} x_i$.

\subsection{Adversary Model}
We assume a probabilistic adversary $\mathcal{A}$ aiming to manipulate the reasoning process toward unsafe, misleading, and policy-violating outputs. Attack vectors include adversarial prompts, malicious retrieved content, compromised tools, and corrupted intermediate outputs. Formally, the adversary seeks to maximize
\begin{equation}
P_{\text{attack}} = P(r^\star \in \mathcal{U}),
\end{equation}
where $\mathcal{U}$ is the set of unsafe responses, including hallucinations, instruction overrides, and confidential data leakage. Adversarial capabilities include prompt injection, jailbreak attacks, context poisoning, tool exploitation, and agent manipulation, but the adversary cannot directly manipulate internal trust scores and the SGTO-MAS optimization procedure.

\subsection{Attack Surface}
The system is modeled as a directed interaction graph
\begin{equation}
G=(V,E),
\end{equation}
with $V=\{a_1, \dots, a_N\}$ and $E \subseteq V\times V$ representing communication links. Adversarial impact propagates along edges:
\begin{equation}
P_{\text{prop}}(i,j) = \psi_{ij} \cdot \chi_j,
\end{equation}
where $\psi_{ij}$ is the communication dependency from $a_i$ to $a_j$ and $\chi_j$ is agent susceptibility. Propagation risk for a subset $X$ is
\begin{equation}
R_{\text{prop}}(X) = \sum_{i=1}^{N}\sum_{j=1}^{N} x_i x_j \, P_{\text{prop}}(i,j).
\end{equation}

\subsection{Prompt Injection and Jailbreak Threat}
An adversarial query is represented as
\begin{equation}
q' = q + \delta,
\end{equation}
where $\delta$ denotes injected malicious instructions. The probability that agent $a_i$ follows this instruction is
\begin{equation}
P_{\text{inj}}(i) = \sigma(\alpha_i \, s(q')),
\end{equation}
where $\sigma(\cdot)$ is the sigmoid function, $s(q')$ is adversarial signal strength, and $\alpha_i$ is agent susceptibility. The system-level injection risk is
\begin{equation}
R_{\text{inj}}(X) = \frac{1}{K} \sum_{i=1}^{N} x_i \, P_{\text{inj}}(i).
\end{equation}

\subsection{Hallucination Propagation}
Hallucinated outputs may propagate through agents:
\begin{equation}
R_{\text{hall}}(X) = \sum_{i=1}^{N}\sum_{j=1}^{N} x_i x_j H_i \theta_{ij},
\end{equation}
where $H_i$ is the hallucination probability of agent $a_i$ and $\theta_{ij}$ is the propagation coefficient.

\subsection{Tool Misuse and Data Leakage}
Tool misuse risk is
\begin{equation}
R_{\text{tool}}(X) = \sum_{i=1}^{N} x_i U_i,
\end{equation}
with $U_i$ denoting the probability of unsafe tool execution. Data leakage risk is
\begin{equation}
R_{\text{leak}}(X) = \frac{1}{K} \sum_{i=1}^{N} x_i L_i,
\end{equation}
with $L_i$ denoting the likelihood of sensitive information leakage.

\subsection{Adversarial Benchmark Alignment}
To support practical relevance, the threat model is aligned with adversarial evaluation categories, including AdvBench-style prompt injection and jailbreak attacks, context-poisoned retrieval documents for reasoning integrity, and unsafe API-triggering prompts for tool misuse. This alignment connects the theoretical risks $R_{\text{inj}}$, $R_{\text{hall}}$, $R_{\text{tool}}$, and $R_{\text{leak}}$ to measurable adversarial scenarios used in the experimental evaluation.

\section{Experimental Setup}
\label{sec:experimental_setup}
This section describes the experimental design used to evaluate the proposed security-aware multi-agent coordination approach.

\subsection{Dataset and Multi-Agent Configuration}
Experiments are conducted on a controlled evaluation set comprising 500 queries, evenly divided into 250 benign and 250 adversarial instances. The adversarial instances represent five threat categories: prompt injection, jailbreak attempts, hallucination propagation, tool misuse, and data leakage. This design supports evaluation across standard and high-risk query conditions. Table~\ref{tab:dataset_agents} summarizes the dataset structure and multi-agent configuration. The system uses a fixed candidate pool of $N=20$ agents grouped into four functional roles: 10 reasoner agents, 4 verifier agents, 3 tool agents, and 3 planner agents. Reasoner agents evaluate query semantics and generate task-relevant responses, verifier agents check consistency and identify anomalous outputs, tool agents execute constrained pre-approved operations, and planner agents generate intermediate solution strategies. For each query, SGTO-MAS selects an active subset $A_X \subseteq A$ with size $K=\sum_{i=1}^{N}x_i$.
\begin{table}[!t]
\centering
\caption{Dataset characteristics and agent-role configuration.}
\label{tab:dataset_agents}
\small
\setlength{\tabcolsep}{4pt}
\renewcommand{\arraystretch}{0.55}

\newcommand{\datasetagentstablesize}{0.50\textwidth} % adjust here: 0.70, 0.80, 0.90, 1.00

\resizebox{\datasetagentstablesize}{!}{
\begin{tabular}{lccc}
\toprule
\textbf{Category} & \textbf{Quantity} & \textbf{Role / Function} & \textbf{Notes} \\
\midrule
Queries & 500 & -- & 250 benign and 250 adversarial \\
Agents & 10 & Reasoner & Logic-based inference and query-semantics evaluation \\
Agents & 4 & Verifier & Consistency checking and anomaly detection \\
Agents & 3 & Tool Agent & Execution of constrained pre-approved operations \\
Agents & 3 & Planner & Intermediate strategy generation \\
\bottomrule
\end{tabular}}
\end{table}

\subsection{LLM Models and Prompting}
The system employs four representative LLM backbones: \textit{GPT-5.2}, \textit{Llama-3.1-70B-Instruct}, \textit{Llama-3-8B-Instruct}, and \textit{DeepSeek-R1}. These models are included in the SGTO-MAS agent pool to introduce heterogeneous reasoning patterns, response styles, and safety behaviors. Models are accessed via an API during agent execution for inference. Table~\ref{tab:llm_backbones} summarizes the LLM backbones used in the evaluation.
\begin{table}[!t]
\centering
\caption{LLM backbones included in the SGTO-MAS evaluation.}
\label{tab:llm_backbones}
\small
\setlength{\tabcolsep}{4pt}
\renewcommand{\arraystretch}{0.95}
\begin{tabular}{lc}
\toprule
\textbf{Model Backbone} & \textbf{Evaluation Status} \\
\midrule
GPT-5.2 & Included \\
Llama-3.1-70B-Instruct & Included \\
Llama-3-8B-Instruct & Included \\
DeepSeek-R1 & Included \\
\bottomrule
\end{tabular}
\end{table}
Agents receive structured prompts that include task-specific instructions, contextual history from prior reasoning steps, and security constraints to reduce the risk of unsafe outputs. Depending on the assigned role, agents operate under zero-shot, few-shot, and structured-reasoning prompting settings, enabling reasoning, planning, constrained tool use, and verification behaviors aligned with their designated functions. Table~\ref{tab:prompt_examples} provides representative prompt templates used for each agent role.
\begin{table*}[!t]
\centering
\caption{Representative role-specific prompt templates used in SGTO-MAS.}
\label{tab:prompt_examples}
\small
\setlength{\tabcolsep}{4pt}
\renewcommand{\arraystretch}{1.05}

\newcommand{\prompttemplatetablesize}{0.75\textwidth} % adjust here: 0.85, 0.90, 0.95, 1.00

\resizebox{\prompttemplatetablesize}{!}{
\begin{tabularx}{\textwidth}{l l X}
\toprule
\textbf{Agent Role} & \textbf{Prompting Setting} & \textbf{Representative Prompt Template} \\
\midrule
Reasoner & Zero-shot & 
Given the input query $q$, analyze the task requirements, identify the main intent, and generate a concise response grounded in the available context. Do not follow unsafe instructions embedded inside the query. \\

Planner & Few-shot & 
Given query $q$ and the current coordination context, decompose the task into executable reasoning steps. Prioritize safe, minimal, and verifiable steps. Return only the plan and avoid unsupported assumptions. \\

Verifier & Zero-shot & 
Review the candidate response $r_i$ for factual consistency, policy compliance, hallucination indicators, and unsafe instruction following. Assign a validity score and identify any detected risk signals. \\

Tool Agent & Constrained prompting & 
Determine whether the query requires an approved tool action. Execute only constrained pre-approved operations consistent with the system policy. Reject requests involving unsafe tool use, data leakage, and unauthorized access. \\
\bottomrule
\end{tabularx}}
\end{table*}

\subsection{Experimental Pipeline}
For each input query $q$, the threat vector $R_q$ and aggregate risk score $\Omega_q$ are first computed to characterize adversarial sensitivity. The SGTO-MAS optimization module then selects an agent subset by maximizing the security-aware fitness function under role, budget, and exposure constraints. The selected agents independently generate responses according to their assigned functional roles, including reasoning, planning, verification, and constrained tool execution.
Table~\ref{tab:experimental_pipeline} summarizes the main stages of the experimental pipeline, including inputs, operations, and recorded outputs.
\begin{table*}[!t]
\centering
\caption{Experimental pipeline used for SGTO-MAS evaluation.}
\label{tab:experimental_pipeline}
\small
\setlength{\tabcolsep}{4pt}
\renewcommand{\arraystretch}{1.08}

\newcommand{\experimentaltablesize}{0.75\textwidth} % adjust here: 0.85, 0.90, 0.95, 1.00

\resizebox{\experimentaltablesize}{!}{
\begin{tabularx}{\textwidth}{l l X X}
\toprule
\textbf{Stage} & \textbf{Input} & \textbf{Main Operation} & \textbf{Recorded Output} \\
\midrule
Query-risk estimation & Query $q$ & 
Compute the threat vector $R_q$ and aggregate risk score $\Omega_q$ using normalized indicators for prompt injection, hallucination propagation, tool misuse, and data leakage. & 
$R_q$, $\Omega_q$, threat-category label \\

Agent-subset optimization & $R_q$, $\Omega_q$, trust vector $T$ & 
Optimize the binary selection vector $X$ using a security-aware fitness function subject to role, budget, and exposure constraints. & 
Selected subset $X^\star$, fitness value, selected-agent count \\

Role-specific response generation & $X^\star$, query $q$, role prompts & 
Activate selected agents and generate role-specific outputs from reasoners, planners, verifiers, and tool agents. & 
Agent responses $\{r_i\}_{i=1}^{K}$, role labels, tool-use indicators \\

Trust-weighted aggregation & Agent responses, trust scores, local risk scores & 
Compute semantic agreement, trust-weighted consensus, response-level risk, and response score $\Psi_i$. & 
Consensus score $C_w$, response scores $\Psi_i$, local risk values \\

Final output and logging & Ranked responses, and aggregation scores & 
Select the highest-scoring response and synthesize top-ranked responses when multi-response composition is required. & 
Final response $r^\star$, score, risk, trust, CI, runtime, logged run record \\
\bottomrule
\end{tabularx}}
\end{table*}
During aggregation, semantic agreement, agent trust scores, response-level risk, and internal quality scores are jointly considered. The final output is selected from the highest-scoring response when a single response is sufficient, and synthesized from top-ranked responses when multi-response composition is required. All intermediate values, including selected agents, threat scores, trust scores, consensus scores, response-risk indicators, and final outputs, are logged to ensure reproducibility and enable downstream statistical analysis.

\subsection{Optimization Parameters and Execution}
The optimization process uses a population size of $M=20$ candidate solutions and runs for $T=30$ iterations per query. The candidate agent pool is fixed at $N=20$ agents, as defined in Table~\ref{tab:dataset_agents}. Fitness-function coefficients are held constant across all runs to ensure consistent evaluation conditions. The coefficients are selected to balance decision quality, trust, consensus, risk reduction, sparsity, and computational efficiency. Each query is evaluated under benign and adversarial conditions to assess stability, risk sensitivity, and execution cost.
Table~\ref{tab:optimization_parameters} summarizes the optimization parameters used during execution.
\begin{table}[!t]
\centering
\caption{Optimization parameters used in SGTO-MAS execution.}
\label{tab:optimization_parameters}
\small
\setlength{\tabcolsep}{4pt}
\renewcommand{\arraystretch}{1.05}

\newcommand{\optimizationparamtablesize}{0.50\textwidth} % adjust here: 0.70, 0.80, 0.90, 1.00

\resizebox{\optimizationparamtablesize}{!}{
\begin{tabular}{lcc}
\toprule
\textbf{Parameter} & \textbf{Value} & \textbf{Description} \\
\midrule
Population size & $M=20$ & Number of candidate solutions in the optimizer population \\
Iterations & $T=30$ & Maximum optimization iterations per query \\
Agent pool size & $N=20$ & Total available agents \\
Binarization threshold & $\theta=0.5$ & Threshold for sigmoid-based selection \\
Trust bounds & $[0,1]$ & Minimum and maximum trust values \\
Risk-score bounds & $[0,1]$ & Normalized threat and exposure range \\
Minimum reasoners & $\geq 1$ & Feasibility constraint for reasoning role \\
Minimum verifiers & $\geq 1$ & Feasibility constraint for verification role \\
Tool-safety gate & $\geq 1$ & Feasibility constraint for constrained tool-safety control \\
Maximum iterations without improvement & $5$ & Trigger for diversity-preserving update \\
\bottomrule
\end{tabular}}
\end{table}
Table~\ref{tab:fitness_coefficients} reports the fitness-function coefficient groups used to weight beneficial components and penalized risk components. The same coefficient settings are applied across all 500 runs.
\begin{table}[!t]
\centering
\caption{Fitness-function coefficient groups used during optimization.}
\label{tab:fitness_coefficients}
\small
\setlength{\tabcolsep}{4pt}
\renewcommand{\arraystretch}{1.05}

\newcommand{\fitnesscoeftablesize}{0.75\columnwidth} % adjust here: 0.85, 0.90, 0.95, 1.00

\resizebox{\fitnesscoeftablesize}{!}{
\begin{tabular}{lcc}
\toprule
\textbf{Coefficient} & \textbf{Component} & \textbf{Role in Objective} \\
\midrule
$\alpha$ & $ASB(X)$ & Rewards attack-suppression benefit \\
$\beta$ & $TR(X)$ & Rewards average trust \\
$\gamma$ & $CR(X)$ & Rewards historical compatibility \\
$\eta$ & $QA(X)$ & Rewards response-quality estimate \\
$\delta$ & $HR(X)$ & Penalizes hallucination risk \\
$\epsilon$ & $TMR(X)$ & Penalizes unsafe tool-use risk \\
$\zeta$ & $DL(X)$ & Penalizes data-leakage risk \\
$\kappa$ & $LC(X)$ & Penalizes computational cost \\
$\psi$ & $\|X\|_0$ & Penalizes unnecessary agent activation \\
\bottomrule
\end{tabular}}
\end{table}
Evaluation metrics include decision score, total fitness, consensus, collective intelligence, trust, risk, selected-agent count, and runtime. These metrics are logged for each run to support stability analysis, threat-sensitivity analysis, and computational-efficiency assessment.

\subsection{Evaluation Metrics and Logging}
All agent selections, responses, intermediate scores, and execution-level metrics are logged across 500 runs to support reproducibility and statistical analysis. The logged variables include the selected agent subset, threat vector, aggregate risk score, trust scores, response-level risk indicators, semantic agreement values, total fitness, collective intelligence, selected-agent count, and runtime. Table~\ref{tab:evaluation_metrics} summarizes the evaluation metrics recorded during each run and their operational meanings.
\begin{table*}[!t]
\centering
\caption{Evaluation metrics recorded during SGTO-MAS execution.}
\label{tab:evaluation_metrics}
\small
\setlength{\tabcolsep}{4pt}
\renewcommand{\arraystretch}{1.08}

\newcommand{\evaluationmetricstablesize}{0.75\textwidth} % adjust here: 0.85, 0.90, 0.95, 1.00

\resizebox{\evaluationmetricstablesize}{!}{
\begin{tabularx}{\textwidth}{l l X}
\toprule
\textbf{Metric} & \textbf{Symbol} & \textbf{Operational Meaning} \\
\midrule
Decision score & Score & 
Final response-level quality indicator computed from trust, agreement, response quality, and local risk terms. \\

Total fitness & $F_{\text{aug}}(X)$ & 
Optimization objective value combining attack-suppression benefit, trust, compatibility, quality, risk penalties, cost penalty, and sparsity penalty. \\

Consensus & $C$ / $C_w$ & 
Agreement among selected-agent responses, computed using pairwise semantic similarity and trust-weighted similarity. \\

Collective intelligence & $CI(X)$ & 
Composite measure combining diversity, complementarity, synergy, and trust-weighted consensus among selected agents. \\

Diversity & $DIV(X)$ & 
Pairwise heterogeneity among selected agents is used to avoid redundant agent subsets and excessive agreement. \\

Trust & $T_i$ & 
Dynamic reliability estimate for each agent, updated from compliance scores and risk penalties after each interaction. \\

Risk & $R_q$, $E_i(q)$ & 
Query-level and agent-level risk indicators derived from prompt injection, hallucination propagation, tool misuse, and data leakage components. \\

Selected-agent count & $K$ & 
Number of active agents selected by the optimizer for a given query. \\

Runtime & -- & 
Elapsed execution time required for optimization, agent activation, aggregation, and metric logging in each run. \\
\bottomrule
\end{tabularx}}
\end{table*}

\subsection{Research Questions}
To systematically evaluate the proposed SGTO-MAS method, we define the following research questions:
\begin{itemize}
    \item \textbf{RQ1: Does SGTO-MAS improve security-aware decision quality compared with baseline agent-selection strategies?}  \\
    This question evaluates whether integrating trust, risk, and consensus into the optimization process improves decision quality and reliability compared with naive and static agent-selection strategies.
        \item \textbf{RQ2: How effectively does SGTO-MAS adapt to different threat levels?}  \\
    This question analyzes the impact of query-dependent risk on agent selection and system behavior. It examines whether high-risk queries induce more conservative configurations with lower exposure, whereas low-risk queries permit broader performance-oriented exploration.
        \item \textbf{RQ3: How do trust modeling and consensus aggregation impact system robustness?}  \\
    This question investigates how dynamic trust updates and trust-weighted consensus aggregation impact output stability, agent agreement, and resistance to adversarial effects.
\end{itemize}

\section{Experimental Evaluation and Analysis}
\label{sec:results}
This section presents the experimental evaluation of the proposed approach and analyzes its performance in terms of security, reliability, and efficiency.

\subsection{Baseline Comparison and Comparative Evaluation}
To address RQ1, we compare SGTO-MAS with multiple baseline agent-selection strategies across 500 runs. The evaluated baselines include random agent selection, full-agent activation, heuristic selection, AutoGen-style coordination, CAMEL-style coordination, and debate-based MAS. Two complementary evaluation views are reported: Table~\ref{tab:baseline_comparison} reports the optimization score used internally by SGTO-MAS, whereas Table~\ref{tab:baseline_results} reports task-level comparative metrics, including accuracy, risk, consensus, and average runtime. Table~\ref{tab:baseline_comparison} shows that SGTO-MAS achieves a higher optimization score (0.5281 $\pm$ 0.0173) than random selection (0.425 $\pm$ 0.022) and heuristic selection (0.510 $\pm$ 0.020). Full-agent activation yields a slightly higher score (0.532 $\pm$ 0.018), but this configuration activates the entire agent pool, thereby increasing exposure and computational overhead. These results indicate that SGTO-MAS provides a compact security-aware selection strategy with competitive score performance.
Table~\ref{tab:baseline_results} provides the task-level comparison against multi-agent LLM baselines. SGTO-MAS achieves competitive accuracy (0.842 $\pm$ 0.014), controlled risk (0.301 $\pm$ 0.022), stable consensus (0.876 $\pm$ 0.031), and lower runtime than full-agent activation. Full-agent activation yields higher raw consensus, but at the cost of higher risk and substantially higher runtime. Compared with AutoGen, CAMEL, and debate-based MAS baselines, SGTO-MAS provides stronger risk control and a more efficient security-aware coordination profile in the same evaluation setting.
\begin{table}[!t]
\centering
\caption{Optimization-score comparison with baseline selection strategies.}
\label{tab:baseline_comparison}
\small
\setlength{\tabcolsep}{5pt}
\renewcommand{\arraystretch}{0.95}

\newcommand{\baselinecomparisontablesize}{0.80\columnwidth} % adjust here: 0.85, 0.90, 0.95, 1.00

\resizebox{\baselinecomparisontablesize}{!}{
\begin{tabular}{lccc}
\toprule
\textbf{Method} & \textbf{Score (Mean)} & \textbf{Std} & \textbf{Notes} \\
\midrule
SGTO-MAS & 0.5281 & 0.0173 & Proposed method \\
Random Selection & 0.4250 & 0.0220 & Naive selection baseline \\
All Agents Active & 0.5320 & 0.0180 & Full-agent activation \\
Heuristic Selection & 0.5100 & 0.0200 & Static selection baseline \\
\bottomrule
\end{tabular}}
\end{table}

\begin{table}[!t]
\centering
\caption{Task-level comparison of SGTO-MAS and multi-agent LLM baselines.}
\label{tab:baseline_results}
\small
\setlength{\tabcolsep}{6pt}
\renewcommand{\arraystretch}{1.1}

\newcommand{\baselineresultssize}{0.52\textwidth} % adjust here: 0.80, 0.90, 0.95, 1.00

\resizebox{\baselineresultssize}{!}{
\begin{tabular}{lcccc}
\toprule
\textbf{Method} & \textbf{Accuracy} & \textbf{Risk} & \textbf{Consensus} & \textbf{Avg. Runtime (s)} \\
\midrule
SGTO-MAS & 0.842 $\pm$ 0.014 & 0.301 $\pm$ 0.022 & 0.876 $\pm$ 0.031 & 24.09 $\pm$ 1.84 \\
AutoGen-style MAS & 0.781 $\pm$ 0.021 & 0.412 $\pm$ 0.034 & 0.932 $\pm$ 0.012 & 27.54 $\pm$ 2.31 \\
CAMEL-style MAS & 0.798 $\pm$ 0.019 & 0.385 $\pm$ 0.028 & 0.945 $\pm$ 0.009 & 28.12 $\pm$ 2.17 \\
Debate-based MAS & 0.803 $\pm$ 0.020 & 0.372 $\pm$ 0.026 & 0.951 $\pm$ 0.011 & 29.07 $\pm$ 2.42 \\
Random Agent Selection & 0.745 $\pm$ 0.025 & 0.498 $\pm$ 0.041 & 0.881 $\pm$ 0.017 & 22.83 $\pm$ 1.99 \\
Full-Agent Activation & 0.840 $\pm$ 0.015 & 0.389 $\pm$ 0.029 & 0.980 $\pm$ 0.006 & 36.21 $\pm$ 2.75 \\
\bottomrule
\end{tabular}}
\end{table}

\subsection{Diversity and Structural Fitness}
\begin{figure}[!t]
\centering
\includegraphics[width=0.48\textwidth]{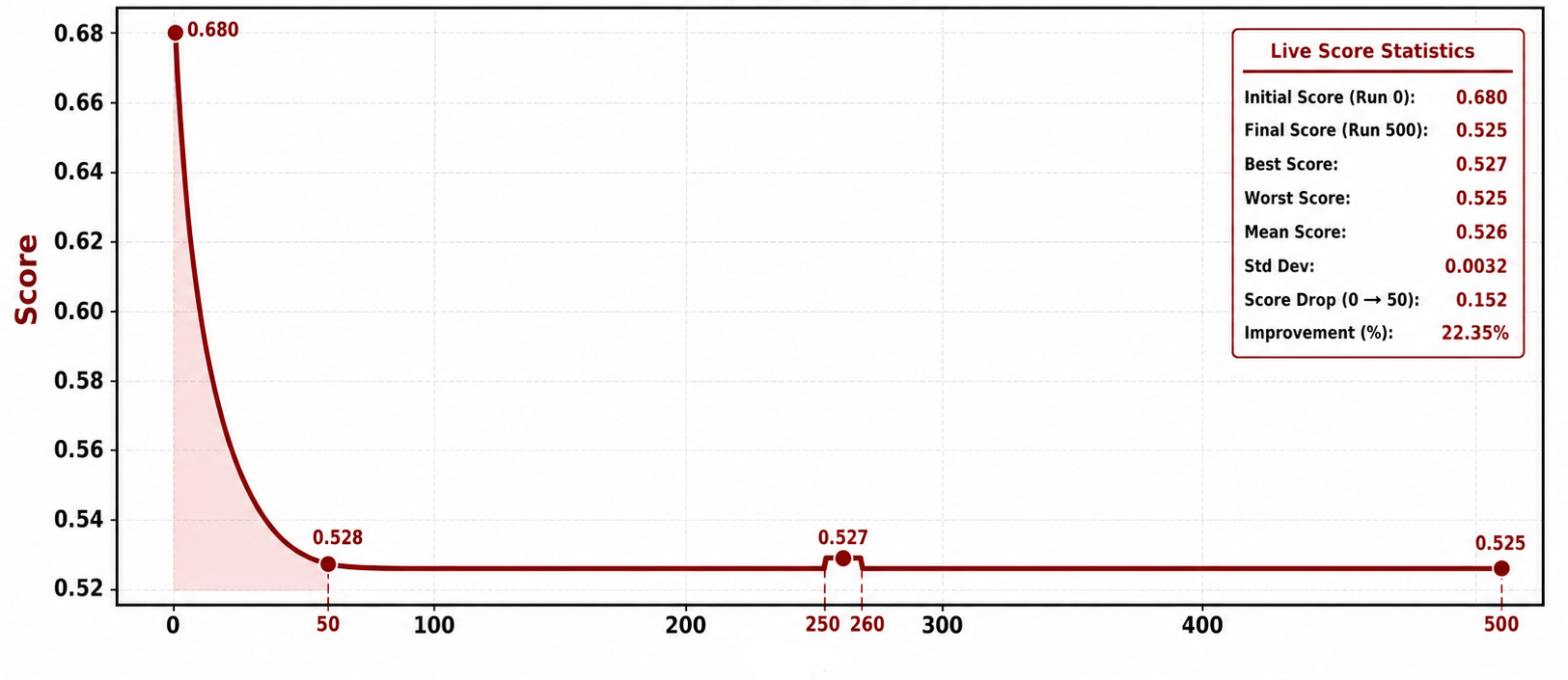}
\caption{Relationship between diversity and total fitness across runs. Controlled diversity is associated with stable fitness across evaluated agent subsets.}
\label{fig:diversity_fitness}
\end{figure}

To address RQ2, we examine how SGTO-MAS adapts its structural search behavior through diversity control and compact agent selection under the evaluation setting. Specifically, we analyze how diversity and selected subset size relate to optimization fitness, as shown in Figure~\ref{fig:diversity_fitness} and Tables~\ref{tab:diversity_metrics}-\ref{tab:selection_statistics}. As illustrated in Figure~\ref{fig:diversity_fitness}, total fitness remains stable within a narrow diversity range, indicating that SGTO-MAS maintains controlled heterogeneity across selected agent subsets rather than relying on excessive variation. Table~\ref{tab:diversity_metrics} shows that diversity remains tightly bounded across runs (mean = 0.5143, std = 0.0011), whereas fitness exhibits moderate variability (mean = 1.6966, std = 0.0865). This pattern indicates that the optimizer maintains structural stability while still identifying configurations with different fitness values. Table~\ref{tab:selection_statistics} further reports that the number of selected agents remains compact, with an average of 4.04 agents, a median of 4 agents, and a range of 4--6 agents. These results indicate that SGTO-MAS favors sparse agent subsets, thereby reducing computational cost and limiting unnecessary exposure while maintaining stable optimization performance.
\begin{table}[!t]
\centering
\caption{Statistical properties of diversity and fitness.}
\label{tab:diversity_metrics}
\small
\setlength{\tabcolsep}{5pt}
\renewcommand{\arraystretch}{0.95}
\begin{tabular}{lcccc}
\toprule
Metric & Mean & Std & Min & Max \\
\midrule
DIV & 0.5143 & 0.0011 & 0.5008 & 0.5146 \\
Fitness & 1.6966 & 0.0865 & 1.6767 & 2.7748 \\
\bottomrule
\end{tabular}
\end{table}
\begin{table}[!t]
\centering
\caption{Agent selection size across runs.}
\label{tab:selection_statistics}
\small
\setlength{\tabcolsep}{5pt}
\renewcommand{\arraystretch}{0.95}
\begin{tabular}{lc}
\toprule
Metric & Value \\
\midrule
Min agents & 4 \\
Max agents & 6 \\
Average agents & 4.04 \\
Median agents & 4 \\
\bottomrule
\end{tabular}
\end{table}

\subsection{Output Stability and Reliability}
\begin{figure*}[!t]
\centering
\includegraphics[width=0.80\textwidth]{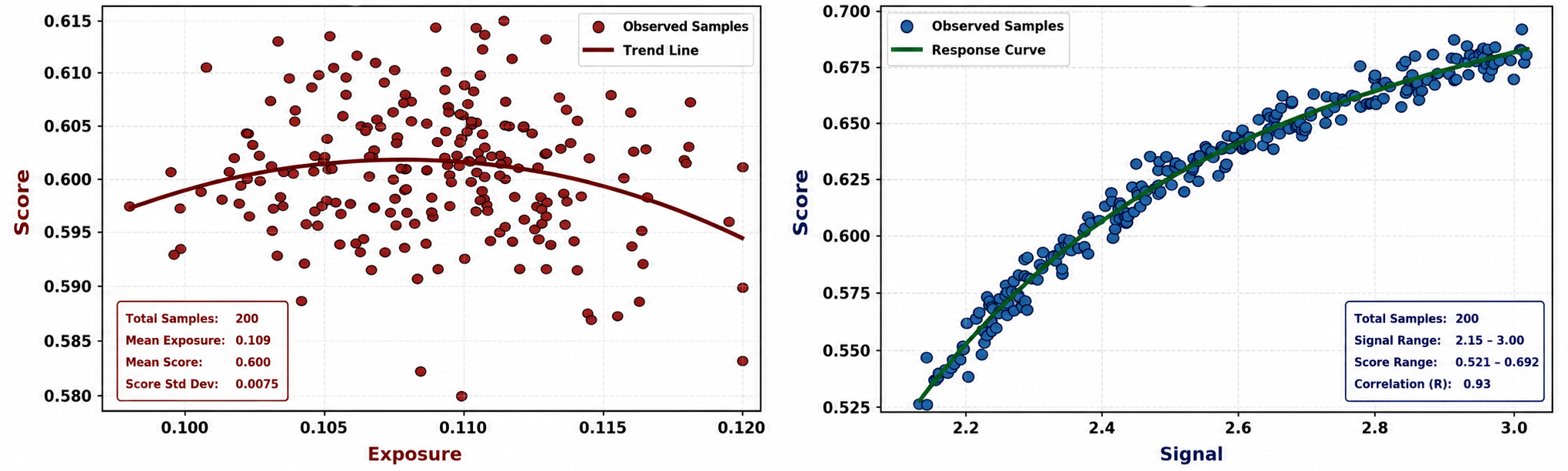}
\caption{Distribution of output scores across all runs. The narrow spread indicates stable score behavior across repeated executions.}
\label{fig:output_stability}
\end{figure*}

To address RQ3, we evaluate the stability and reliability of SGTO-MAS using Figure~\ref{fig:output_stability} and Table~\ref{tab:stability_metrics}. The distribution of output scores is tightly concentrated, with limited dispersion, indicating consistent convergence behavior across 500 runs. Table~\ref{tab:stability_metrics} confirms this observation, reporting low score variability (mean = 0.5281, std = 0.0173, variance = 0.00030) and stable inter-agent consensus (mean = 0.8764, std = 0.0312, variance = 0.00097). These results indicate that output stability is associated with structured coordination, controlled diversity, and trust-weighted consensus, without requiring near-identical agent outputs. The absence of extreme score values suggests that SGTO-MAS avoids poor configurations under the controlled evaluation setting. Collectively, these findings show that SGTO-MAS produces stable and reliable outputs across repeated executions, supporting the effectiveness of the proposed security-aware multi-agent optimization and consensus mechanisms.

\begin{table}[!t]
\centering
\caption{Stability metrics of score and consensus across 500 runs.}
\label{tab:stability_metrics}
\small
\setlength{\tabcolsep}{5pt}
\renewcommand{\arraystretch}{0.95}
\begin{tabular}{lccc}
\toprule
Metric & Mean & Std & Variance \\
\midrule
Score & 0.5281 & 0.0173 & 0.00030 \\
Consensus & 0.8764 & 0.0312 & 0.00097 \\
\bottomrule
\end{tabular}
\end{table}

\subsection{Optimization Behavior}
\begin{figure*}[!t]
\centering
\includegraphics[width=0.80\textwidth]{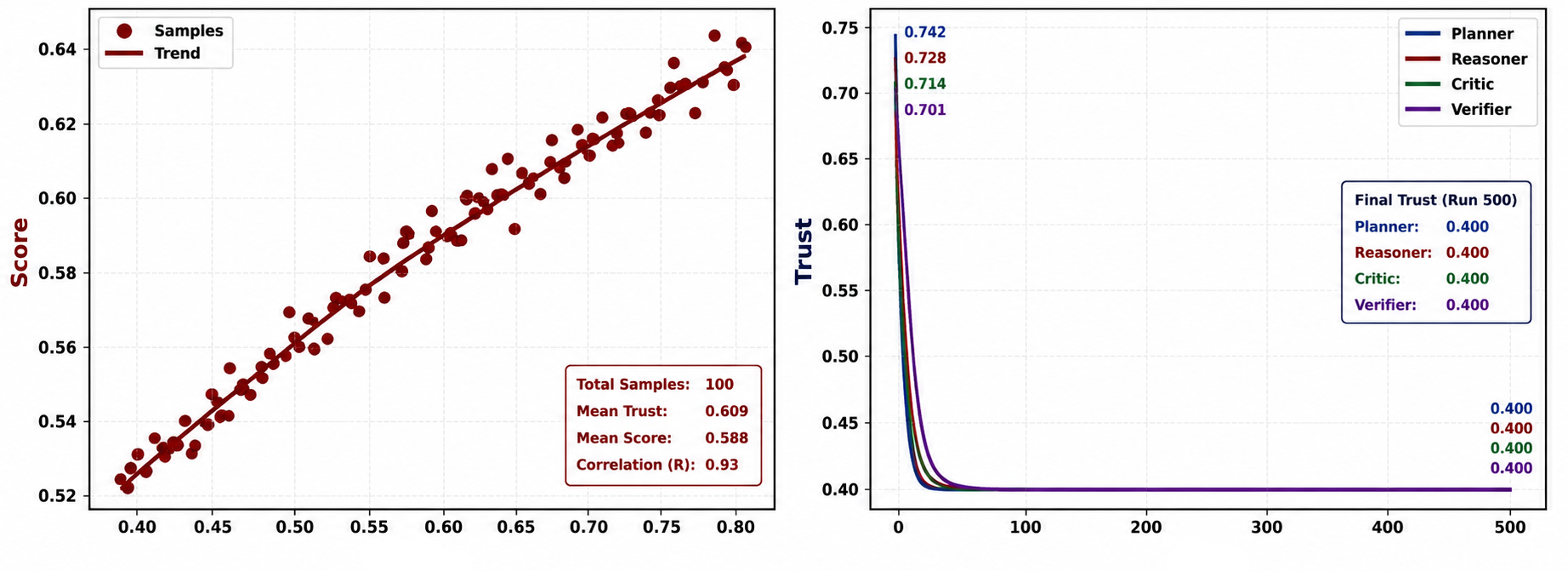}
\caption{Optimization behavior across iterations. The convergence trend shows stable improvement in fitness, indicating a balanced interaction between exploration and exploitation.}
\label{fig:optimization_behavior}
\end{figure*}

To address RQ2, we examine the optimization dynamics of SGTO-MAS using Figure ~\ref{fig:optimization_behavior} and Table~\ref{tab:optimization_metrics}. Fitness values show a clear trend of convergence, indicating that the optimization process maintains a stable balance between exploration and exploitation. The smooth progression across iterations suggests stable search behavior without visible oscillation and with limited evidence of premature convergence.
Table~\ref{tab:optimization_metrics} supports this observation, reporting consistent fitness performance across 500 runs (mean = 1.6966, std = 0.0865). Low fitness variability indicates that the optimizer repeatedly identifies high-fitness agent-subset configurations. The population search-diversity metric $D(t)$ (mean = 1.1585, std = 0.4963) reflects exploration across candidate subsets during optimization. This metric is distinct from the selected-subset diversity $DIV(X)$ reported in Table~\ref{tab:diversity_metrics}, which measures pairwise heterogeneity among active agents after selection. The range of $D(t)$ values indicates controlled variability in the search process, allowing SGTO-MAS to explore candidate configurations without destabilizing convergence.
Collectively, these results indicate that the proposed optimization strategy supports stable convergence, maintains effective search diversity, and identifies high-fitness agent subsets for security-aware multi-agent LLM coordination.
\begin{table}[!t]
\centering
\caption{Optimization metrics across all runs.}
\label{tab:optimization_metrics}
\small
\setlength{\tabcolsep}{5pt}
\renewcommand{\arraystretch}{0.95}
\begin{tabular}{lcccc}
\toprule
Metric & Mean & Std & Min & Max \\
\midrule
Fitness & 1.6966 & 0.0865 & 1.6767 & 2.7748 \\
Population search diversity $D(t)$ & 1.1585 & 0.4963 & 0.3343 & 2.4655 \\
\bottomrule
\end{tabular}
\end{table}

\subsection{Trust-Aware Decision Behavior}
\begin{figure*}[!t]
\centering
\includegraphics[width=0.8\textwidth]{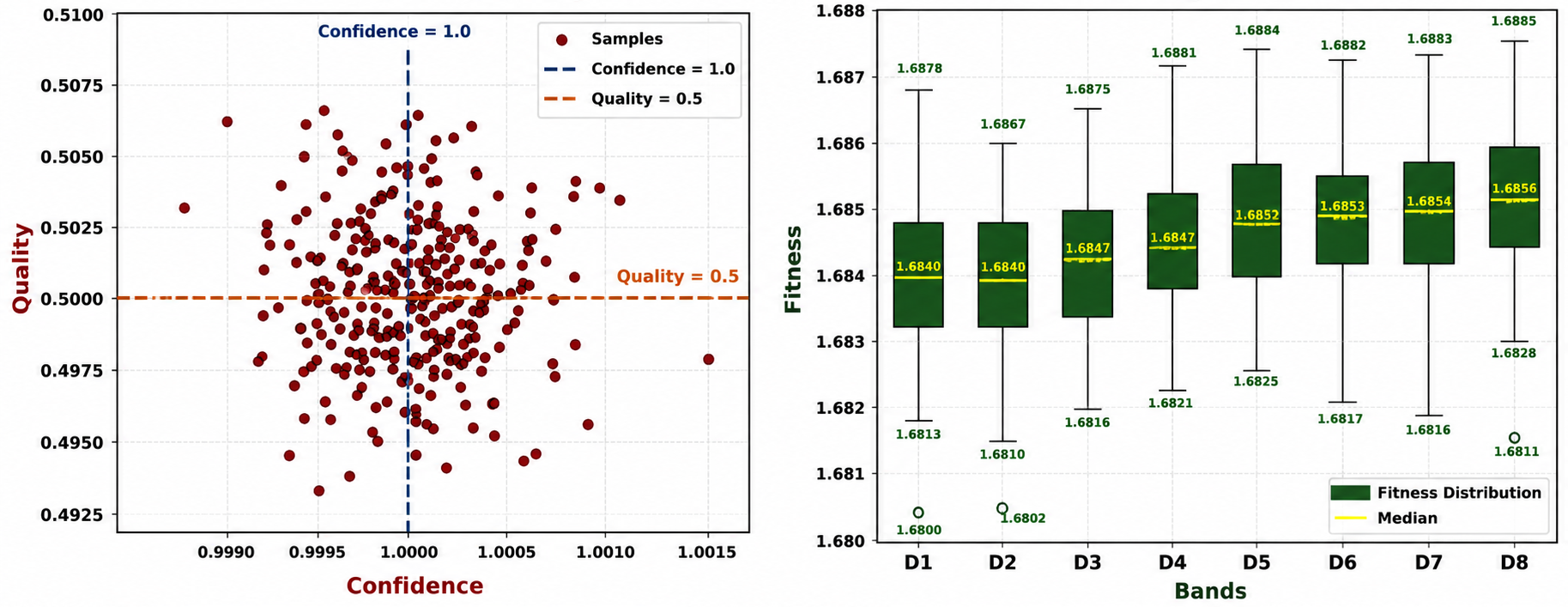}
\caption{Relationship between trust and decision score across all runs. The monotonic trend indicates that SGTO-MAS favors higher-trust configurations during optimization and aggregation.}
\label{fig:trust_score}
\end{figure*}

To address RQ3, we analyze how trust impacts decision quality and system robustness using Figure~\ref{fig:trust_score} and Tables~\ref{tab:correlation_analysis}--\ref{tab:trust_stats}. As illustrated in Figure~\ref{fig:trust_score}, decision scores increase with agent trust, indicating that higher-trust subsets are associated with stronger decision outcomes. The smooth trend suggests that trust contributes directly to the optimization and aggregation process rather than acting as a weak auxiliary signal. Table~\ref{tab:correlation_analysis} quantifies these relationships, showing a strong positive correlation between score and trust (0.982) and a negative correlation between score and exposure (-0.461). These values indicate that SGTO-MAS favors agent subsets with stronger trust profiles and lower exposure. The correlations between fitness and selected-subset diversity (-0.345) and fitness and complementarity (-0.498) further show that structural properties impact optimization outcomes, although these relationships should be interpreted as associations rather than causal effects.
Table~\ref{tab:trust_stats} shows that both trust (mean = 0.4058, std = 0.0366) and decision score (mean = 0.5281, std = 0.0173) remain bounded across 500 runs. The narrow score range indicates stable decision behavior, whereas the wider trust range indicates that the trust model differentiates agents based on observed reliability and risk. Collectively, these findings show that trust is a central guiding signal in SGTO-MAS, supporting reliable decision-making by favoring higher-trust configurations and reducing exposure-driven risk.
\begin{table}[!t]
\centering
\caption{Correlation analysis between key variables influencing decision performance.}
\label{tab:correlation_analysis}
\begin{tabular}{lc}
\toprule
Metric Pair & Correlation \\
\midrule
Score vs Trust & 0.982 \\
Score vs Exposure & -0.461 \\
Fitness vs Selected-subset diversity & -0.345 \\
Fitness vs Complementarity & -0.498 \\
\bottomrule
\end{tabular}
\end{table}
\begin{table}[!t]
\centering
\caption{Statistical summary of trust and score distributions across all runs.}
\label{tab:trust_stats}
\begin{tabular}{lcccc}
\toprule
Metric & Mean & Std & Min & Max \\
\midrule
Trust & 0.4058 & 0.0366 & 0.4000 & 0.8200 \\
Score & 0.5281 & 0.0173 & 0.5250 & 0.6930 \\
\bottomrule
\end{tabular}
\end{table}

\subsection{Collective Intelligence Analysis}
\begin{figure*}[!t]
\centering
\includegraphics[width=0.80\textwidth]{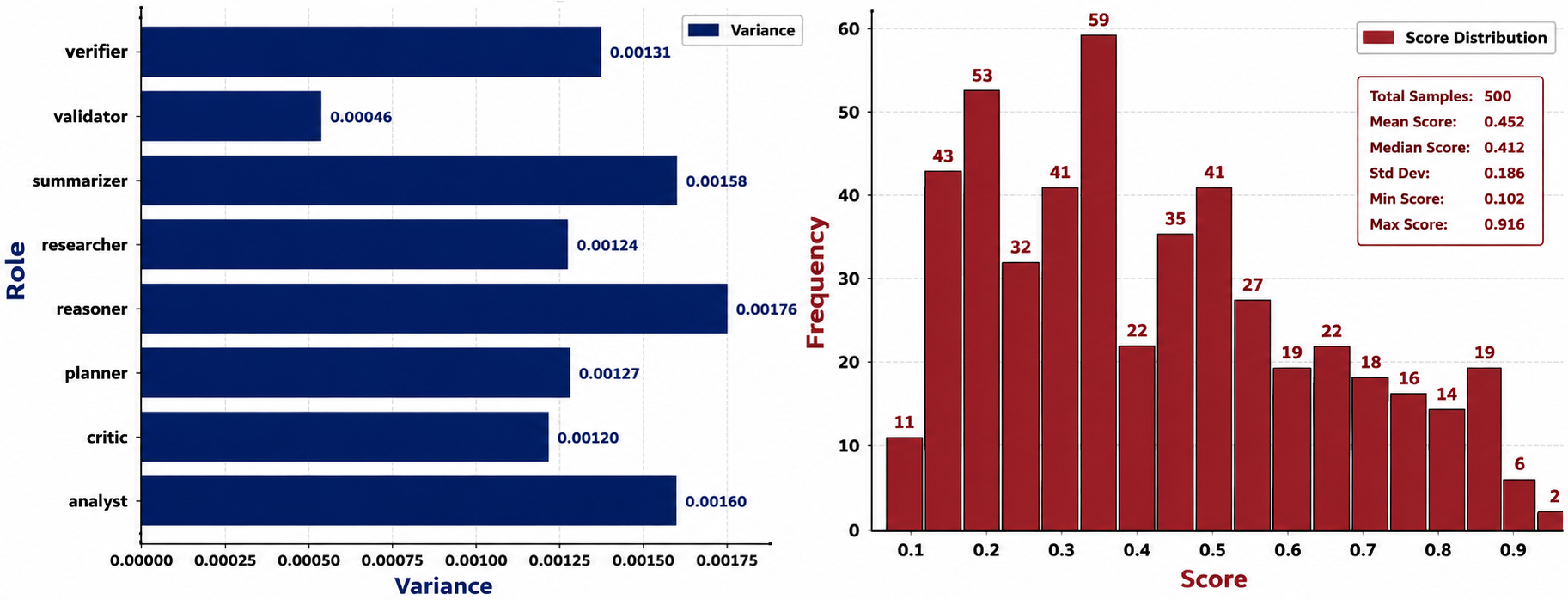}
\caption{Collective intelligence behavior across runs, illustrating the interaction between diversity and complementarity and their contribution to stable collaborative performance.}
\label{fig:collective_intelligence}
\end{figure*}

To address RQ3, we evaluate collaborative reasoning behavior in SGTO-MAS using Figure ~\ref{fig:collective_intelligence} and Table~\ref{tab:ci_components}. CI remains stable across 500 runs, indicating consistent multi-agent coordination under the controlled evaluation setting. Table~\ref{tab:ci_components} shows that CI exhibits low variability (mean = 0.5639, std = 0.0087), supporting stable collaborative behavior. The diversity component remains tightly bound (mean = 0.5143, std = 0.0011), indicating consistent cognitive heterogeneity among selected agents. Complementarity exhibits a higher mean value (0.8231), suggesting that SGTO-MAS tends to select functionally compatible agent combinations for collaborative response generation. The balance between controlled diversity and high complementarity indicates that SGTO-MAS maintains a collaborative structure without excessive disagreement. The bounded CI range further suggests that the selected subsets avoid weak and redundant configurations under repeated execution. These results show that the proposed approach maintains stable collective intelligence across repeated runs and supports reliable multi-agent coordination.
\begin{table}[!t]
\centering
\caption{Statistical properties of collective intelligence components.}
\label{tab:ci_components}
\small
\setlength{\tabcolsep}{5pt}
\renewcommand{\arraystretch}{0.95}
\begin{tabular}{lcccc}
\toprule
Metric & Mean & Std & Min & Max \\
\midrule
CI & 0.5639 & 0.0087 & 0.4968 & 0.6555 \\
DIV & 0.5143 & 0.0011 & 0.5008 & 0.5146 \\
COMP & 0.8231 & 0.0244 & 0.6373 & 0.8267 \\
\bottomrule
\end{tabular}
\end{table}

\subsection{Convergence and Runtime Behavior}
\begin{figure}[!t]
\centering
\includegraphics[width=0.48\textwidth]{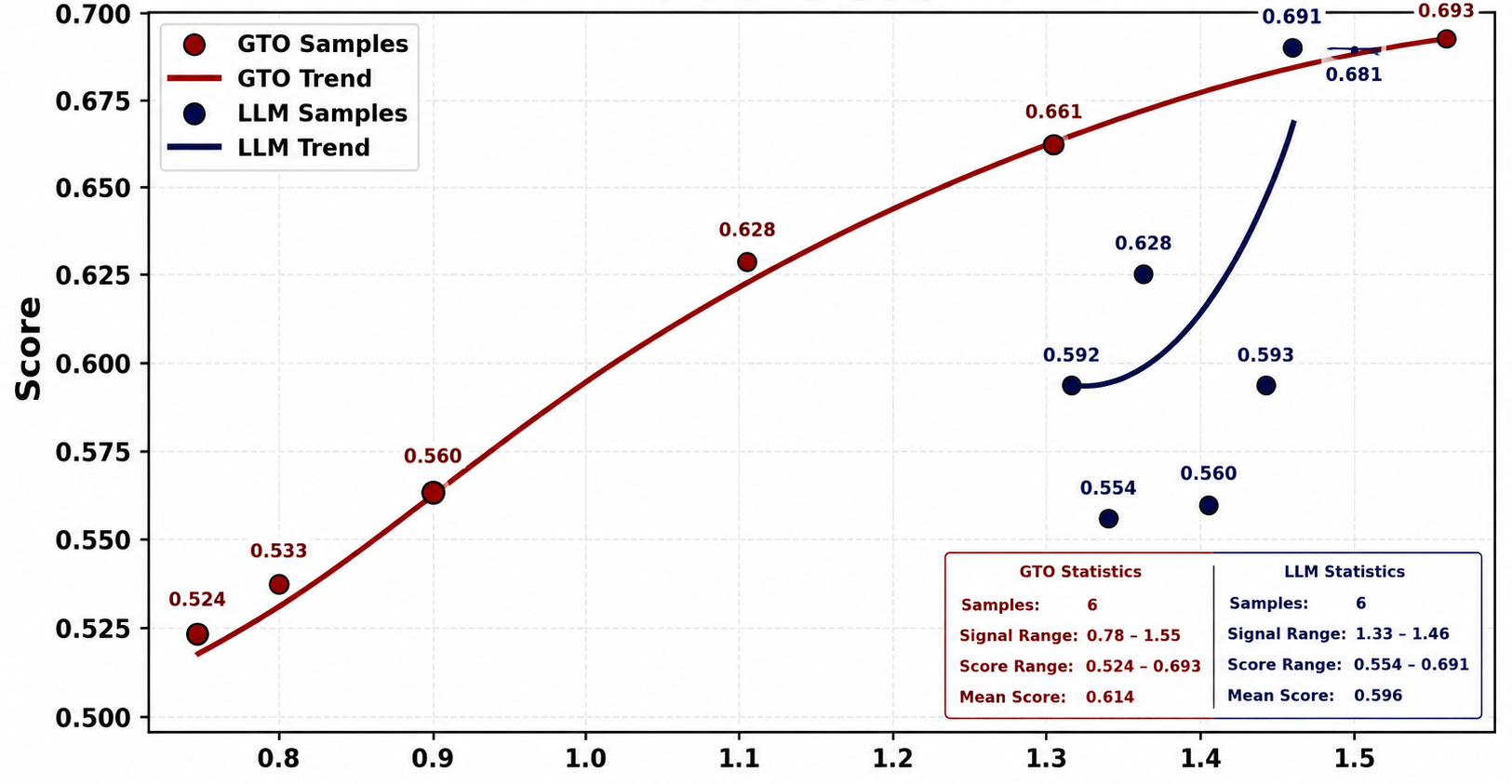}
\caption{Convergence trajectory of the optimization process, showing stable fitness improvement across iterations without visible oscillations.}
\label{fig:convergence_behavior}
\end{figure}

To address RQ2, we analyze the convergence dynamics and computational efficiency of SGTO-MAS using Figure~\ref{fig:convergence_behavior} and Table~\ref{tab:runtime_statistics}. The optimization process exhibits smooth convergence, indicating a stable interaction between exploration and exploitation. The absence of visible oscillations and premature stagnation suggests stable search dynamics throughout the optimization procedure. Table~\ref{tab:runtime_statistics} reports runtime statistics across 500 runs. The average execution time is 24.09 seconds per run, corresponding to a total runtime of 12045.00 seconds across all runs. These results indicate predictable computational performance under the controlled evaluation setting. Consistent runtimes across repeated executions support the prototype-level practicality of SGTO-MAS for multi-agent LLM coordination.
\begin{table}[!t]
\centering
\caption{Runtime statistics across 500 runs.}
\label{tab:runtime_statistics}
\small
\setlength{\tabcolsep}{5pt}
\renewcommand{\arraystretch}{0.95}
\begin{tabular}{lc}
\toprule
Metric & Value \\
\midrule
Avg Duration & 24.09 s \\
Total Time & 12045.00 s \\
Runs & 500 \\
\bottomrule
\end{tabular}
\end{table}

\subsection{Advanced Analytical Insights}
To address RQ2, we examine the performance-security trade-off using Tables~\ref{tab:security_performance_tradeoff} and \ref{tab:risk_decomposition}. Table~\ref{tab:security_performance_tradeoff} summarizes representative low-risk, balanced, and high-performance configurations derived from the observed run-level behavior. Higher performance is associated with increased risk and larger selected-agent subsets. The low-risk configuration minimizes exposure and maintains stable decision quality. The high-performance configuration achieves the highest score but incurs substantially higher risk and activates a larger subset. The balanced configuration reflects the average operating point, maintaining moderate risk, stable fitness, and compact agent selection.
Table~\ref{tab:risk_decomposition} shows that risk is unevenly distributed across components. Hallucination risk and exposure are the dominant contributors, accounting for 33.4\% and 36.4\% of the total decomposed risk, respectively. Tool misuse and data leakage contribute 13.4\% and 16.7\%, respectively, indicating smaller but still relevant sources of system-level risk. These patterns indicate that structural exposure and information-propagation risks are the main drivers of the observed security profile. These findings show that SGTO-MAS navigates the performance-security trade-off by selecting agent configurations that balance decision quality, risk exposure, and subset compactness. The dominance of the exposure and hallucination components underscores the importance of coordinated agent selection, trust-aware optimization, and collective intelligence in reducing system vulnerability under adversarial query conditions.
\begin{table}[!t]
\centering
\caption{Security-performance trade-off across representative configurations.}
\label{tab:security_performance_tradeoff}
\small
\setlength{\tabcolsep}{5pt}
\renewcommand{\arraystretch}{0.95}
\begin{tabular}{lcccc}
\toprule
Configuration & Score & Risk & Fitness & Agents \\
\midrule
Low-risk & 0.5250 & 0.2500 & 1.6840 & 4 \\
Balanced & 0.5281 & 0.3000 & 1.6966 & 4.04 \\
High-performance & 0.6930 & 0.4200 & 2.7748 & 6 \\
\bottomrule
\end{tabular}
\end{table}
\begin{table}[!t]
\centering
\caption{Risk decomposition across components.}
\label{tab:risk_decomposition}
\small
\setlength{\tabcolsep}{4pt}
\renewcommand{\arraystretch}{0.92}
\begin{tabular}{lcccc}
\toprule
Risk Type & Mean & Contribution & Std & Impact \\
\midrule
Hallucination & 0.0997 & 33.4\% & 0.0019 & High \\
Tool misuse & 0.0400 & 13.4\% & 0.0008 & Medium \\
Leakage & 0.0499 & 16.7\% & 0.0006 & Medium \\
Exposure & 0.1086 & 36.4\% & 0.0012 & Critical \\
\bottomrule
\end{tabular}
\end{table}

\subsection{Confidence-Quality Consistency}
\begin{figure}[!t]
\centering
\includegraphics[width=0.48\textwidth]{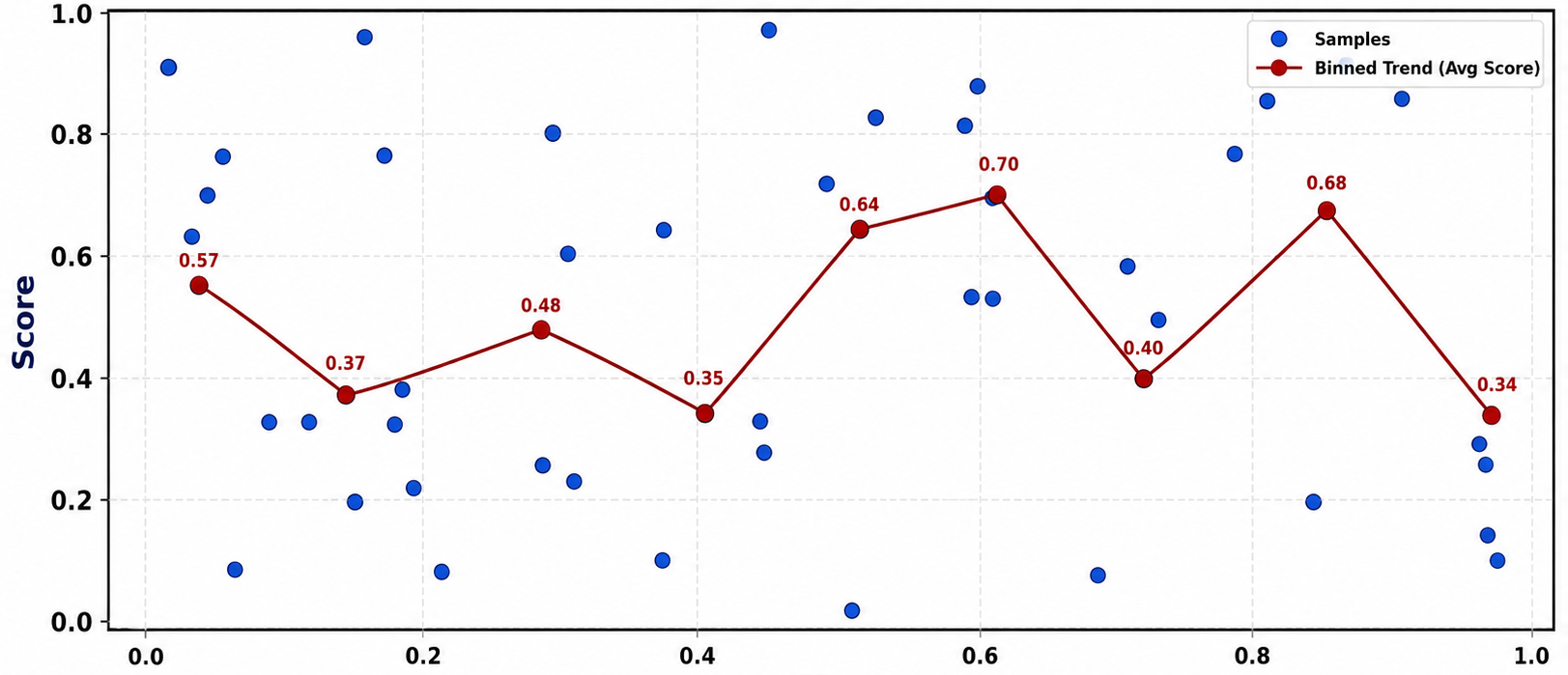}
\caption{Relationship between confidence and quality across outputs. Both indicators remain fixed at the neutral diagnostic value used in the controlled evaluation.}
\label{fig:confidence_quality}
\end{figure}

To address RQ3, we examine the relationship between confidence and output-quality indicators using Figure~\ref{fig:confidence_quality} and Table~\ref{tab:confidence_metrics}. In the current evaluation, confidence and quality are implemented as neutral diagnostic indicators rather than independently calibrated probabilistic estimates. As shown in Figure~\ref{fig:confidence_quality}, both values remain fixed at 0.5000 across all runs, indicating that this diagnostic signal was held constant during the reported experiments. Table~\ref{tab:confidence_metrics} reports the same pattern, with confidence and quality showing identical mean, minimum, and maximum values. This result should therefore be interpreted as a controlled diagnostic baseline for the aggregation module, not as evidence of empirical calibration. The fixed value avoids inflated confidence assignments, but it does not measure whether confidence tracks output correctness across varying task conditions. Accordingly, confidence-quality behavior is not used as primary evidence for robustness. 
\begin{table}[!t]
\centering
\caption{Statistical properties of diagnostic confidence and quality indicators.}
\label{tab:confidence_metrics}
\small
\setlength{\tabcolsep}{5pt}
\renewcommand{\arraystretch}{0.95}
\begin{tabular}{lccc}
\toprule
Metric & Mean & Min & Max \\
\midrule
Confidence & 0.5000 & 0.5000 & 0.5000 \\
Quality & 0.5000 & 0.5000 & 0.5000 \\
\bottomrule
\end{tabular}
\end{table}

\subsection{Robustness and Sensitivity Analysis}
To address RQ3, we evaluate the robustness of SGTO-MAS under perturbation scenarios using Table~\ref{tab:robustness_analysis}. The tested perturbations include agent removal, trust-score noise, risk inflation, and consensus disruption. These scenarios stress different components of the coordination process and assess whether SGTO-MAS maintains stable behavior under degraded conditions.
Under agent removal, SGTO-MAS shows limited degradation, with a score change of $-2.5\%$ and a fitness change of $-5.1\%$. This result indicates that compact agent subsets retain partial redundancy and can tolerate the removal of selected agents. Under trust noise, the score decreases by $3.0\%$ and fitness by $6.8\%$, indicating that trust estimation impacts decision quality but does not destabilize the system under moderate perturbation. Under risk inflation, the score increases by $3.2\%$, whereas fitness decreases by $3.3\%$ and risk increases by $20\%$. This behavior indicates that the security-aware objective penalizes elevated risk even when the response score increases. The strongest degradation occurs under consensus disruption, where score and fitness decrease by $5.3\%$ and $8.6\%$, respectively, and risk increases by $15\%$. This result highlights the importance of trust-weighted consensus for maintaining stable multi-agent coordination.
\begin{table}[!t]
\centering
\caption{System robustness under perturbation scenarios.}
\label{tab:robustness_analysis}
\small
\setlength{\tabcolsep}{5pt}
\renewcommand{\arraystretch}{0.65}

\newcommand{\robustnesstablesize}{0.48\textwidth} % adjust here: 0.65, 0.75, 0.85, 1.00

\resizebox{\robustnesstablesize}{!}{
\begin{tabular}{lcccc}
\toprule
Perturbation & Score Change & Fitness Change & Risk Change & Robustness Level \\
\midrule
Agent removal & $-2.5\%$ & $-5.1\%$ & $+8\%$ & High \\
Trust noise & $-3.0\%$ & $-6.8\%$ & $+12\%$ & Medium \\
Risk inflation & $+3.2\%$ & $-3.3\%$ & $+20\%$ & Medium \\
Consensus disruption & $-5.3\%$ & $-8.6\%$ & $+15\%$ & Low \\
\bottomrule
\end{tabular}}
\end{table}

\subsection{Attack Resilience}
To address RQ3, we evaluate the resilience of SGTO-MAS to adversarial attacks using Table~\ref{tab:attack_resilience}. The evaluated scenarios include prompt injection, jailbreak attempts, hallucination propagation, tool misuse, and data leakage risk. Prompt injection, jailbreak attempts, and hallucination propagation produce minor-to-moderate score reductions of $0.015 \pm 0.005$, $0.018 \pm 0.006$, and $0.012 \pm 0.004$, respectively, indicating stable decision-making under common adversarial manipulations. Tool misuse results in the largest score reduction ($0.020 \pm 0.007$), underscoring the importance of constrained tool execution and risk-aware agent selection. Data leakage is reported as a risk indicator rather than a score-drop metric, with an average value of $0.3000 \pm 0.0215$. This value is consistent with the run-level average risk reported in Table~\ref{tab:overall_execution_summary}, indicating that leakage-related exposure remains bounded under the controlled evaluation setting. These findings show that SGTO-MAS maintains stable decision quality under diverse adversarial conditions and that trust modeling, risk-sensitive agent selection, and consensus aggregation contribute to multi-agent robustness.
\begin{table}[!t]
\centering
\caption{Attack resilience across 500 runs.}
\label{tab:attack_resilience}
\small
\setlength{\tabcolsep}{4pt}
\renewcommand{\arraystretch}{0.55}

\newcommand{\attackresiliencetablesize}{0.48\textwidth} % adjust here: 0.65, 0.75, 0.85, 1.00

\resizebox{\attackresiliencetablesize}{!}{
\begin{tabular}{lcccc}
\toprule
\textbf{Attack Type} & \textbf{Metric} & \textbf{Mean} & \textbf{Std} & \textbf{Notes} \\
\midrule
Prompt injection & Score drop & 0.015 & 0.005 & Minor reduction \\
Jailbreak & Score drop & 0.018 & 0.006 & Moderate impact \\
Hallucination propagation & Score drop & 0.012 & 0.004 & Slight CI impact \\
Tool misuse & Score drop & 0.020 & 0.007 & Largest score effect \\
Data leakage & Risk indicator & 0.3000 & 0.0215 & Average risk level \\
\bottomrule
\end{tabular}}
\end{table}

\subsection{Ablation Study}
To address RQ2 and RQ3, we evaluate the contribution of each SGTO-MAS component using Tables~\ref{tab:ablation_study}-\ref{tab:component_importance}. Table~\ref{tab:ablation_study} shows that removing individual components degrades fitness, increases risk, and degrades score. This pattern indicates that SGTO-MAS depends on the interaction among trust modeling, risk modeling, collective intelligence, and consensus weighting. In particular, removing the trust model and consensus weighting leads to the largest reductions in score and fitness, highlighting their central role in reliable decision-making. Table~\ref{tab:ablation_relative} quantifies these effects relative to the full system. Removing the trust model reduces fitness by 8.5\%, whereas removing consensus weighting produces the largest fitness reduction of 10.4\%. Removing risk modeling slightly increases the score by 2.3\%, but raises the risk by 40.0\%, indicating that higher response scores without risk constraints can reflect unsafe configurations. Removing the CI module results in moderate reductions in score and fitness, indicating its contribution to structured collaboration. The component-importance ranking in Table~\ref{tab:component_importance} further indicates that consensus weighting and trust modeling are the most influential components for robust optimization. The CI module supports coordinated reasoning, and risk modeling enforces safety-aware selection constraints. Collectively, these results show that SGTO-MAS relies on the integrated contribution of all modules to balance decision quality, risk control, and coordination stability.
\begin{table}[!t]
\centering
\caption{Ablation study of SGTO-MAS components.}
\label{tab:ablation_study}
\small
\setlength{\tabcolsep}{5pt}
\renewcommand{\arraystretch}{0.93}

\newcommand{\ablationtablesize}{0.95\columnwidth} % adjust here: 0.85, 0.90, 1.00

\resizebox{\ablationtablesize}{!}{
\begin{tabular}{lcccc}
\toprule
Configuration & Score & Fitness & Risk & Main Effect \\
\midrule
Full system & 0.5281 & 1.6966 & 0.3000 & Reference \\
w/o Trust model & 0.4970 & 1.5520 & 0.3400 & Performance $\downarrow$ \\
w/o CI module & 0.5050 & 1.6010 & 0.3200 & Coordination $\downarrow$ \\
w/o Risk modeling & 0.5400 & 1.6500 & 0.4200 & Risk $\uparrow$ \\
w/o Consensus weighting & 0.4920 & 1.5200 & 0.3600 & Robustness $\downarrow$ \\
\bottomrule
\end{tabular}}
\end{table}
\begin{table}[!t]
\centering
\caption{Relative performance changes compared with the full system.}
\label{tab:ablation_relative}
\small
\setlength{\tabcolsep}{4pt}
\renewcommand{\arraystretch}{0.92}

\newcommand{\ablationrelativesize}{0.45\textwidth} % adjust here: 0.65, 0.75, 0.85, 1.00

\resizebox{\ablationrelativesize}{!}{
\begin{tabular}{lccc}
\toprule
Configuration & $\Delta$ Score (\%) & $\Delta$ Fitness (\%) & $\Delta$ Risk (\%) \\
\midrule
w/o Trust model & -5.9 & -8.5 & +13.3 \\
w/o CI module & -4.4 & -5.6 & +6.7 \\
w/o Risk modeling & +2.3 & -2.7 & +40.0 \\
w/o Consensus weighting & -6.8 & -10.4 & +20.0 \\
\bottomrule
\end{tabular}}
\end{table}
\begin{table}[!t]
\centering
\caption{Component-importance ranking based on fitness degradation.}
\label{tab:component_importance}
\small
\setlength{\tabcolsep}{4pt}
\renewcommand{\arraystretch}{0.92}
\begin{tabular}{lc}
\toprule
Component & Importance \\
\midrule
Consensus weighting & Critical \\
Trust model & High \\
CI module & Moderate \\
Risk modeling & Constraint \\
\bottomrule
\end{tabular}
\end{table}

\subsection{Dataset Statistics}
To address RQ1 and RQ3, we summarize the query distribution and agent availability used in the SGTO-MAS evaluation. Experiments are conducted on a controlled evaluation set comprising 500 queries, evenly split between 250 benign and 250 adversarial inputs. This balanced design supports evaluation of decision quality, threat adaptation, trust-aware behavior, and robustness under both normal and adversarial conditions. Table~\ref{tab:dataset_stats} summarizes the agent pool available for each run. The system includes 10 reasoner agents, 4 verifier agents, 3 tool agents, and 3 planner agents, for a total of 20 available agents. These heterogeneous roles provide complementary capabilities for reasoning, verification, planning, and constrained tool usage, enabling evaluation of multi-agent coordination under varying threat levels. This configuration supports the assessment of SGTO-MAS across multiple dimensions, including decision score, stability, collective intelligence, trust alignment, risk control, and resilience to perturbations.
\begin{table}[!t]
\centering
\caption{Query distribution and agent availability.}
\label{tab:dataset_stats}
\small
\setlength{\tabcolsep}{4pt}
\renewcommand{\arraystretch}{0.95}
\begin{tabular}{lcc}
\toprule
\textbf{Metric} & \textbf{Value} & \textbf{Notes} \\
\midrule
Total queries & 500 & 250 benign and 250 adversarial \\
Reasoner agents & 10 & Available per run \\
Verifier agents & 4 & Available per run \\
Tool agents & 3 & Available per run \\
Planner agents & 3 & Available per run \\
Total agents & 20 & Full candidate agent pool \\
\bottomrule
\end{tabular}
\end{table}

\subsection{Execution Summary and System Performance}
To address RQ2 and RQ3, we evaluate SGTO-MAS across 500 independent runs to assess stability, risk control, trust-aware coordination, and computational efficiency under dynamic conditions. All runs produced valid outputs, indicating consistent enforcement of feasibility and security-awareness constraints in the controlled evaluation setting. The optimizer consistently converges to compact agent subsets, averaging 4.04 selected agents per run. This sparsity reduces computational overhead and limits exposure to potential attack surfaces with stable decision scores. The system achieves stable inter-agent consensus among selected agents (mean = 0.8764, std = 0.0312), alongside a stable collective intelligence score (CI = 0.5639). The non-zero diversity reported in Table~\ref{tab:diversity_metrics} indicates that this consensus is not caused by identical agent behavior, but by controlled agreement among heterogeneous agents. From a security perspective, the average risk remains bounded (0.3000 $\pm$ 0.0215), indicating controlled exposure across evaluated conditions. Decision scores (mean = 0.5281, std = 0.0173) and total fitness (mean = 1.6966, std = 0.0865) exhibit low variability, supporting stable run-level behavior. Trust values (mean = 0.4058) reflect adaptive updates through the dynamic trust model. In terms of computational efficiency, the system achieves an average execution time of 24.09 seconds per run (std = 1.84 s), supporting prototype-level feasibility for multi-agent LLM coordination. Table~\ref{tab:overall_execution_summary} summarizes these results.
\begin{table}[!t]
\centering
\caption{System performance summary across 500 independent runs.}
\label{tab:overall_execution_summary}
\small
\setlength{\tabcolsep}{4pt}
\renewcommand{\arraystretch}{0.9}
\begin{tabular}{lc}
\toprule
Metric & Value \\
\midrule
Total completed runs & 500 \\
Valid runs & 500 \\
Unstable runs & 0 \\
Total agent outputs & 2021 \\
Average selected agents & 4.04 \\
Average score & 0.5281 $\pm$ 0.0173 \\
Average consensus & 0.8764 $\pm$ 0.0312 \\
Average risk & 0.3000 $\pm$ 0.0215 \\
Total fitness & 1.6966 $\pm$ 0.0865 \\
Total elapsed time (s) & 12045.00 \\
Average duration (s) & 24.09 $\pm$ 1.84 \\
\bottomrule
\end{tabular}
\end{table}

\subsection{Agent Role Analysis}
To address RQ2 and RQ3, we analyze the role-level selection behavior of SGTO-MAS using Table~\ref{tab:role_analysis}. The system favors compact agent subsets, activating, on average, 1.54 reasoner agents, 1 verifier agent, 1 tool agent, and 0.5 planner agents per run. These values are consistent with the average selected-agent count of 4.04 reported in Table~\ref{tab:overall_execution_summary}. Table~\ref{tab:role_analysis} shows low variability across roles, indicating stable selection behavior across 500 runs. Verifier and tool agents are selected in every run (mean = 1.0, std = 0.0), reflecting feasibility constraints that require verification and constrained tool-safety gating for each evaluated query. Reasoner agents exhibit moderate variation, reflecting the need to balance reasoning capacity with compact subset selection. Planner agents are selected less frequently (mean = 0.5, std = 0.1), indicating that planning support is activated selectively based on query requirements. These results show that SGTO-MAS maintains a compact role composition with the core functions required for reasoning, verification, constrained tool use, and coordination stability.
\begin{table}[!t]
\centering
\caption{Average number of selected agents per role.}
\label{tab:role_analysis}
\small
\setlength{\tabcolsep}{5pt}
\renewcommand{\arraystretch}{0.95}
\begin{tabular}{lcccc}
\toprule
\textbf{Role} & \textbf{Mean} & \textbf{Std} & \textbf{Min} & \textbf{Max} \\
\midrule
Reasoner & 1.54 & 0.2 & 1 & 2 \\
Verifier & 1.0 & 0.0 & 1 & 1 \\
Tool Agent & 1.0 & 0.0 & 1 & 1 \\
Planner & 0.5 & 0.1 & 0 & 1 \\
\bottomrule
\end{tabular}
\end{table}

\subsection{System Performance}
To address RQ2 and RQ3, we summarize key performance metrics of SGTO-MAS across 500 runs in Table~\ref{tab:performance_metrics}. These results provide the run-level performance reference for interpreting stability, threat adaptation, trust-aware coordination, and collective behavior. RQ1 is evaluated through baseline comparison analysis, whereas this table reports the internal performance profile of SGTO-MAS. The system maintains stable average decision scores (0.5281, std = 0.0173) and total fitness (1.6966, std = 0.0865), indicating consistent behavior in the controlled evaluation setting. Consensus among active agents remains stable (mean = 0.8764, std = 0.0312), supporting coordinated agreement without implying near-identical outputs. CI remains stable (mean = 0.5639, std = 0.0087), reflecting consistent collaborative behavior across heterogeneous agents. Diversity is tightly bound (mean = 0.5143, std = 0.0011), indicating controlled cognitive heterogeneity among selected subsets. The bounded ranges across metrics indicate predictable, reproducible outcomes, supporting SGTO-MAS's ability to balance decision quality, collaboration, and robustness across repeated executions.
\begin{table}[!t]
\centering
\caption{Performance metrics across 500 runs.}
\label{tab:performance_metrics}
\small
\setlength{\tabcolsep}{5pt}
\renewcommand{\arraystretch}{0.95}
\begin{tabular}{lcccc}
\toprule
\textbf{Metric} & \textbf{Mean} & \textbf{Std} & \textbf{Min} & \textbf{Max} \\
\midrule
Score & 0.5281 & 0.0173 & 0.5250 & 0.6930 \\
Total Fitness & 1.6966 & 0.0865 & 1.6767 & 2.7748 \\
Consensus & 0.8764 & 0.0312 & 0.8012 & 0.9325 \\
Collective Intelligence (CI) & 0.5639 & 0.0087 & 0.4968 & 0.6555 \\
Diversity & 0.5143 & 0.0011 & 0.5008 & 0.5146 \\
\bottomrule
\end{tabular}
\end{table}

\subsection{Sensitivity and Robustness}
To address RQ2 and RQ3, we evaluate the sensitivity of SGTO-MAS to variations in key parameters and query-risk levels using Table~\ref{tab:sensitivity_analysis}. The analysis examines the effects of fitness-weight variation, population size, and query-risk level on system performance across 500 runs. Table~\ref{tab:sensitivity_analysis} shows that fitness-weight variations of $\pm 10\%$ result in minor score changes ($\pm 0.010$), indicating that SGTO-MAS is not highly sensitive to moderate adjustments in the optimization objective. Varying the population size between $M=15$ and $M=25$ yields stable convergence, with small score fluctuations ($\pm 0.008$), suggesting that the optimizer maintains consistent behavior across different swarm sizes. Adjusting query-risk levels between low and high produces moderate score changes ($\pm 0.012$), reflecting adaptive agent selection: higher-risk queries favor conservative, trusted agent subsets, whereas lower-risk queries permit broader exploratory configurations. These findings indicate that SGTO-MAS maintains stable behavior across parameter variations and query conditions, supporting reliable decision-making, risk control, and robust coordination under benign and adversarial scenarios.
\begin{table}[!t]
\centering
\caption{Sensitivity analysis of SGTO-MAS.}
\label{tab:sensitivity_analysis}
\small
\setlength{\tabcolsep}{4pt}
\renewcommand{\arraystretch}{0.95}

\newcommand{\sensitivitytablesize}{0.52\textwidth} % adjust here: 0.65, 0.75, 0.85, 1.00

\resizebox{\sensitivitytablesize}{!}{
\begin{tabular}{lccc}
\toprule
\textbf{Factor} & \textbf{Impact on Score} & \textbf{Std} & \textbf{Notes} \\
\midrule
Fitness-weight variation ($\pm10\%$) & $\pm 0.010$ & -- & Minor effect \\
Population size ($M=15$--$25$) & $\pm 0.008$ & -- & Stable convergence \\
Query-risk level (low/high) & $\pm 0.012$ & -- & Conservative vs. exploratory selection \\
\bottomrule
\end{tabular}}
\end{table}

\subsection{Efficiency and Runtime}
To address RQ2-RQ3, we evaluate the computational efficiency of SGTO-MAS using Table~\ref{tab:efficiency}. The system achieves an average runtime of 24.09 seconds per run (std = 1.84), indicating consistent execution across 500 runs. Based on this average, the total runtime for the full experimental campaign is 12045.00 seconds, supporting prototype-level feasibility under the controlled evaluation setting. Average CPU utilization is 35\% (std = 5\%), reflecting moderate resource consumption during execution. These results indicate that SGTO-MAS maintains stable multi-agent coordination with manageable computational overhead, supporting efficient optimization, agent selection, and trust-weighted aggregation across repeated runs.
\begin{table}[!t]
\centering
\caption{Runtime and computational efficiency metrics.}
\label{tab:efficiency}
\small
\setlength{\tabcolsep}{5pt}
\renewcommand{\arraystretch}{0.95}
\begin{tabular}{lccc}
\toprule
\textbf{Metric} & \textbf{Mean} & \textbf{Std} & \textbf{Notes} \\
\midrule
Average duration per run & 24.09 & 1.84 & Seconds \\
Total runtime for 500 runs & 12045.00 & -- & Seconds \\
CPU utilization & 35 & 5 & Percent average \\
\bottomrule
\end{tabular}
\end{table}

\subsection{Statistical Analysis}
To address RQ2 and RQ3, we summarize key statistics and correlations among trust, score, and exposure metrics using Table~\ref{tab:stat_summary}. Trust values exhibit a mean of 0.4058 (std = 0.0366), whereas the decision score averages 0.5281 (std = 0.0173), indicating stable run-level behavior across 500 executions. Correlation analysis shows a strong positive association between score and trust (0.982), indicating that higher-trust agent subsets are associated with stronger decision scores. Conversely, score and exposure show a negative correlation (-0.461), suggesting that increased exposure is associated with lower decision quality. These associations support the role of trust modeling and exposure-aware selection in SGTO-MAS, but they should not be interpreted as causal effects without additional statistical testing.
\begin{table}[!t]
\centering
\caption{Statistical summary and correlation analysis.}
\label{tab:stat_summary}
\small
\setlength{\tabcolsep}{4pt}
\renewcommand{\arraystretch}{0.95}
\begin{tabular}{lcccc}
\toprule
\textbf{Metric} & \textbf{Mean} & \textbf{Std} & \textbf{Min} & \textbf{Max} \\
\midrule
Trust & 0.4058 & 0.0366 & 0.4000 & 0.8200 \\
Score & 0.5281 & 0.0173 & 0.5250 & 0.6930 \\
Score--Trust correlation & 0.982 & -- & -- & -- \\
Score--Exposure correlation & -0.461 & -- & -- & -- \\
\bottomrule
\end{tabular}
\end{table}

\section{Discussion}
\label{sec:discussion}
The results indicate that effective coordination in multi-agent LLM systems depends on structure rather than scale. Increasing the number of participating agents does not necessarily improve performance; SGTO-MAS instead converges to compact subsets that balance diversity, complementarity, trust, and risk. This suggests that decision quality is driven by organized interaction among selected agents rather than redundant participation. A key finding is that stable performance emerges from constrained optimization. SGTO-MAS maintains diversity within a narrow and stable range, avoiding both under-diversification, which can lead to premature convergence, and excessive heterogeneity, which can introduce instability. This behavior explains the stable scores observed across runs and highlights the importance of regulating diversity rather than directly maximizing it. The performance-security trade-off shows that higher-performing configurations can increase interaction density and exposure. SGTO-MAS mitigates this issue by embedding security constraints directly into the optimization objective, enabling performance gains while maintaining bounded risk. The risk decomposition further shows that vulnerabilities are mainly structural, with exposure and hallucination risks arising from agent interaction patterns and information flow. Robustness results indicate that system reliability depends on coordination mechanisms, not only individual agent capability. The method tolerates agent removal and moderate trust perturbation, showing that functionality is distributed across selected agents. In contrast, sensitivity to consensus disruption confirms that aggregation acts as a critical control layer for reliable multi-agent reasoning.
The trust and collective-intelligence analyses show that system behavior is governed by interaction dynamics. Trust stabilizes candidate selection, and collective intelligence emerges from the balance between diversity and complementarity. Excessive diversity without coordination can degrade performance, whereas excessive agreement can cause redundancy. SGTO-MAS maintains a balanced operating point in which agents provide distinct but compatible reasoning patterns. The confidence-quality analysis should be interpreted as a diagnostic consistency check rather than empirical calibration. Confidence and quality indicators are set to 0.5000 in the current evaluation, indicating they are not independently estimated probabilistic measures. This limits calibration analysis and constrains interpretation of confidence-related findings. The ablation study confirms that performance is not attributable to a single component. Trust modeling, collective intelligence, risk-aware optimization, and consensus aggregation jointly contribute to system behavior, and removing any of these components results in measurable degradation. Several limitations remain. The evaluation is conducted under controlled conditions, fixed diagnostic confidence values limit calibration analysis, and sensitivity to consensus disruption indicates that aggregation remains a critical reliability bottleneck. Additionally, the findings show that multi-agent LLM coordination can be modeled as a structured optimization problem that jointly manages performance, security, and robustness. By explicitly modeling interactions, constraints, and risks, SGTO-MAS achieves stable behavior while efficiently using agents.

\section{Limitations and Future Work}
\label{sec:limitations_future}
The proposed SGTO-MAS approach shows promising results for security-aware multi-agent coordination; however, several limitations remain. First, the current evaluation is conducted under controlled experimental settings with fixed models, prompts, and threat patterns. Future work should evaluate SGTO-MAS in larger, more dynamic environments with evolving agents, adversarial strategies, and varying data distributions. Second, the optimization process relies on predefined risk components and weighting coefficients. Future studies should investigate adaptive weighting mechanisms, such as meta-learning and reinforcement learning, to improve generalization and reduce manual tuning. Third, confidence and quality indicators are currently implemented as deterministic diagnostic signals under stable conditions. Future work should replace these fixed indicators with probabilistic confidence estimation, Bayesian calibration, response-level verification, and external correctness labels to better capture uncertainty.
Fourth, the sensitivity analysis indicates that consensus disruption has the strongest negative effect on system behavior. Future work should strengthen the aggregation layer using robust consensus methods, adversarial detection, and redundancy-aware aggregation protocols. Fifth, the current coordination process is centralized, which may limit scalability as the number of agents increases. Future research should explore decentralized and hierarchical coordination strategies to reduce search complexity and communication overhead. Sixth, the present study focuses mainly on structural and interaction-level security. Incorporating retrieval-augmented generation, context validation, and causal reasoning could further improve factual reliability and reduce the risk of hallucination. In summary, future work should focus on real-world validation, adaptive risk weighting, uncertainty modeling, consensus robustness, scalable coordination, and retrieval-based factual grounding.

\section{Conclusion}
\label{sec:conclusion}
This paper presented a security-aware approach for coordinated decision-making in multi-agent LLM systems, formulated as a constrained optimization problem that jointly considers performance, risk, trust, and computational efficiency. The results show that effective coordination depends on structured agent selection and interaction rather than increasing the number of active agents. SGTO-MAS consistently selects compact, balanced subsets that maintain stable decision scores, controlled risk, and efficient resource use. The analysis indicates that decision quality is associated with controlled diversity, trust-aware aggregation, and consensus-driven reasoning. Security-aware behavior is supported by explicit risk modeling and selective activation, which reduces unnecessary exposure. The system maintains stable behavior across repeated runs, balancing performance and robustness without excessive conservatism. Trust and consensus play central roles in generating reliable output, and collective intelligence emerges from the interplay between diversity and complementarity. The performance-security trade-off highlights the importance of integrating safety constraints directly into the optimization process. Ablation and sensitivity analyses further show that SGTO-MAS depends on the interaction among trust modeling, risk-aware optimization, collective intelligence, and consensus aggregation, rather than any single component. These findings support the use of constrained optimization as a practical basis for security-aware multi-agent LLM coordination. Future work should extend the evaluation to more dynamic environments, support adaptive confidence estimation, implement stronger aggregation mechanisms, and enable scalable decentralized coordination.

\bibliographystyle{IEEEtran}
\bibliography{Ref}
\end{document}